\documentclass[11pt,a4paper,english,superscriptaddress,aps,preprint]{revtex4}
\usepackage{amsmath}
\usepackage{amssymb}
\usepackage{graphicx}
\makeatletter
\usepackage{babel}
\newcommand{\bea}{\begin{eqnarray}}
\newcommand{\eea}{\end{eqnarray}}

\newcommand{\pa}{\partial}
\renewcommand{\a}{\alpha}

\renewcommand{\b}{\beta}

\newcommand{\q}{\theta}

\newcommand{\be}{\begin{equation}}
\newcommand{\ee}{\end{equation}}

\begin{document}
\immediate\write16{<<WARNING: LINEDRAW macros work with emTeX-dvivers
                    and other drivers supporting emTeX \special's
                    (dviscr, dvihplj, dvidot, dvips, dviwin, etc.) >>}

\newdimen\Lengthunit       \Lengthunit  = 1.5cm
\newcount\Nhalfperiods     \Nhalfperiods= 9
\newcount\magnitude        \magnitude = 1000

\catcode`\*=11
\newdimen\L*   \newdimen\d*   \newdimen\d**
\newdimen\dm*  \newdimen\dd*  \newdimen\dt*
\newdimen\a*   \newdimen\b*   \newdimen\c*
\newdimen\a**  \newdimen\b**
\newdimen\xL*  \newdimen\yL*
\newdimen\rx*  \newdimen\ry*
\newdimen\tmp* \newdimen\linwid*

\newcount\k*   \newcount\l*   \newcount\m*
\newcount\k**  \newcount\l**  \newcount\m**
\newcount\n*   \newcount\dn*  \newcount\r*
\newcount\N*   \newcount\*one \newcount\*two  \*one=1 \*two=2
\newcount\*ths \*ths=1000
\newcount\angle*  \newcount\q*  \newcount\q**
\newcount\angle** \angle**=0
\newcount\sc*     \sc*=0

\newtoks\cos*  \cos*={1}
\newtoks\sin*  \sin*={0}

\catcode`\[=13

\def\rotate(#1){\advance\angle**#1\angle*=\angle**
\q**=\angle*\ifnum\q**<0\q**=-\q**\fi
\ifnum\q**>360\q*=\angle*\divide\q*360\multiply\q*360\advance\angle*-\q*\fi
\ifnum\angle*<0\advance\angle*360\fi\q**=\angle*\divide\q**90\q**=\q**
\def\sgcos*{+}\def\sgsin*{+}\relax
\ifcase\q**\or
 \def\sgcos*{-}\def\sgsin*{+}\or
 \def\sgcos*{-}\def\sgsin*{-}\or
 \def\sgcos*{+}\def\sgsin*{-}\else\fi
\q*=\q**
\multiply\q*90\advance\angle*-\q*
\ifnum\angle*>45\sc*=1\angle*=-\angle*\advance\angle*90\else\sc*=0\fi
\def[##1,##2]{\ifnum\sc*=0\relax
\edef\cs*{\sgcos*.##1}\edef\sn*{\sgsin*.##2}\ifcase\q**\or
 \edef\cs*{\sgcos*.##2}\edef\sn*{\sgsin*.##1}\or
 \edef\cs*{\sgcos*.##1}\edef\sn*{\sgsin*.##2}\or
 \edef\cs*{\sgcos*.##2}\edef\sn*{\sgsin*.##1}\else\fi\else
\edef\cs*{\sgcos*.##2}\edef\sn*{\sgsin*.##1}\ifcase\q**\or
 \edef\cs*{\sgcos*.##1}\edef\sn*{\sgsin*.##2}\or
 \edef\cs*{\sgcos*.##2}\edef\sn*{\sgsin*.##1}\or
 \edef\cs*{\sgcos*.##1}\edef\sn*{\sgsin*.##2}\else\fi\fi
\cos*={\cs*}\sin*={\sn*}\global\edef\gcos*{\cs*}\global\edef\gsin*{\sn*}}\relax
\ifcase\angle*[9999,0]\or
[999,017]\or[999,034]\or[998,052]\or[997,069]\or[996,087]\or
[994,104]\or[992,121]\or[990,139]\or[987,156]\or[984,173]\or
[981,190]\or[978,207]\or[974,224]\or[970,241]\or[965,258]\or
[961,275]\or[956,292]\or[951,309]\or[945,325]\or[939,342]\or
[933,358]\or[927,374]\or[920,390]\or[913,406]\or[906,422]\or
[898,438]\or[891,453]\or[882,469]\or[874,484]\or[866,499]\or
[857,515]\or[848,529]\or[838,544]\or[829,559]\or[819,573]\or
[809,587]\or[798,601]\or[788,615]\or[777,629]\or[766,642]\or
[754,656]\or[743,669]\or[731,681]\or[719,694]\or[707,707]\or
\else[9999,0]\fi}

\catcode`\[=12

\def\GRAPH(hsize=#1)#2{\hbox to #1\Lengthunit{#2\hss}}

\def\Linewidth#1{\global\linwid*=#1\relax
\global\divide\linwid*10\global\multiply\linwid*\mag
\global\divide\linwid*100\special{em:linewidth \the\linwid*}}

\Linewidth{.4pt}
\def\sm*{\special{em:moveto}}
\def\sl*{\special{em:lineto}}
\let\moveto=\sm*
\let\lineto=\sl*
\newbox\spm*   \newbox\spl*
\setbox\spm*\hbox{\sm*}
\setbox\spl*\hbox{\sl*}

\def\mov#1(#2,#3)#4{\rlap{\L*=#1\Lengthunit
\xL*=#2\L* \yL*=#3\L*
\xL*=\xscale\xL* \yL*=\yscale\yL*
\rx* \the\cos*\xL* \tmp* \the\sin*\yL* \advance\rx*-\tmp*
\ry* \the\cos*\yL* \tmp* \the\sin*\xL* \advance\ry*\tmp*
\kern\rx*\raise\ry*\hbox{#4}}}

\def\rmov*(#1,#2)#3{\rlap{\xL*=#1\yL*=#2\relax
\rx* \the\cos*\xL* \tmp* \the\sin*\yL* \advance\rx*-\tmp*
\ry* \the\cos*\yL* \tmp* \the\sin*\xL* \advance\ry*\tmp*
\kern\rx*\raise\ry*\hbox{#3}}}

\def\lin#1(#2,#3){\rlap{\sm*\mov#1(#2,#3){\sl*}}}

\def\arr*(#1,#2,#3){\rmov*(#1\dd*,#1\dt*){\sm*
\rmov*(#2\dd*,#2\dt*){\rmov*(#3\dt*,-#3\dd*){\sl*}}\sm*
\rmov*(#2\dd*,#2\dt*){\rmov*(-#3\dt*,#3\dd*){\sl*}}}}

\def\arrow#1(#2,#3){\rlap{\lin#1(#2,#3)\mov#1(#2,#3){\relax
\d**=-.012\Lengthunit\dd*=#2\d**\dt*=#3\d**
\arr*(1,10,4)\arr*(3,8,4)\arr*(4.8,4.2,3)}}}

\def\arrlin#1(#2,#3){\rlap{\L*=#1\Lengthunit\L*=.5\L*
\lin#1(#2,#3)\rmov*(#2\L*,#3\L*){\arrow.1(#2,#3)}}}

\def\dasharrow#1(#2,#3){\rlap{{\Lengthunit=0.9\Lengthunit
\dashlin#1(#2,#3)\mov#1(#2,#3){\sm*}}\mov#1(#2,#3){\sl*
\d**=-.012\Lengthunit\dd*=#2\d**\dt*=#3\d**
\arr*(1,10,4)\arr*(3,8,4)\arr*(4.8,4.2,3)}}}

\def\clap#1{\hbox to 0pt{\hss #1\hss}}

\def\ind(#1,#2)#3{\rlap{\L*=.1\Lengthunit
\xL*=#1\L* \yL*=#2\L*
\rx* \the\cos*\xL* \tmp* \the\sin*\yL* \advance\rx*-\tmp*
\ry* \the\cos*\yL* \tmp* \the\sin*\xL* \advance\ry*\tmp*
\kern\rx*\raise\ry*\hbox{\lower2pt\clap{$#3$}}}}

\def\sh*(#1,#2)#3{\rlap{\dm*=\the\n*\d**
\xL*=\xscale\dm* \yL*=\yscale\dm* \xL*=#1\xL* \yL*=#2\yL*
\rx* \the\cos*\xL* \tmp* \the\sin*\yL* \advance\rx*-\tmp*
\ry* \the\cos*\yL* \tmp* \the\sin*\xL* \advance\ry*\tmp*
\kern\rx*\raise\ry*\hbox{#3}}}

\def\calcnum*#1(#2,#3){\a*=1000sp\b*=1000sp\a*=#2\a*\b*=#3\b*
\ifdim\a*<0pt\a*-\a*\fi\ifdim\b*<0pt\b*-\b*\fi
\ifdim\a*>\b*\c*=.96\a*\advance\c*.4\b*
\else\c*=.96\b*\advance\c*.4\a*\fi
\k*\a*\multiply\k*\k*\l*\b*\multiply\l*\l*
\m*\k*\advance\m*\l*\n*\c*\r*\n*\multiply\n*\n*
\dn*\m*\advance\dn*-\n*\divide\dn*2\divide\dn*\r*
\advance\r*\dn*
\c*=\the\Nhalfperiods5sp\c*=#1\c*\ifdim\c*<0pt\c*-\c*\fi
\multiply\c*\r*\N*\c*\divide\N*10000}

\def\dashlin#1(#2,#3){\rlap{\calcnum*#1(#2,#3)\relax
\d**=#1\Lengthunit\ifdim\d**<0pt\d**-\d**\fi
\divide\N*2\multiply\N*2\advance\N*\*one
\divide\d**\N*\sm*\n*\*one\sh*(#2,#3){\sl*}\loop
\advance\n*\*one\sh*(#2,#3){\sm*}\advance\n*\*one
\sh*(#2,#3){\sl*}\ifnum\n*<\N*\repeat}}

\def\dashdotlin#1(#2,#3){\rlap{\calcnum*#1(#2,#3)\relax
\d**=#1\Lengthunit\ifdim\d**<0pt\d**-\d**\fi
\divide\N*2\multiply\N*2\advance\N*1\multiply\N*2\relax
\divide\d**\N*\sm*\n*\*two\sh*(#2,#3){\sl*}\loop
\advance\n*\*one\sh*(#2,#3){\kern-1.48pt\lower.5pt\hbox{\rm.}}\relax
\advance\n*\*one\sh*(#2,#3){\sm*}\advance\n*\*two
\sh*(#2,#3){\sl*}\ifnum\n*<\N*\repeat}}

\def\shl*(#1,#2)#3{\kern#1#3\lower#2#3\hbox{\unhcopy\spl*}}

\def\trianglin#1(#2,#3){\rlap{\toks0={#2}\toks1={#3}\calcnum*#1(#2,#3)\relax
\dd*=.57\Lengthunit\dd*=#1\dd*\divide\dd*\N*
\divide\dd*\*ths \multiply\dd*\magnitude
\d**=#1\Lengthunit\ifdim\d**<0pt\d**-\d**\fi
\multiply\N*2\divide\d**\N*\sm*\n*\*one\loop
\shl**{\dd*}\dd*-\dd*\advance\n*2\relax
\ifnum\n*<\N*\repeat\n*\N*\shl**{0pt}}}

\def\wavelin#1(#2,#3){\rlap{\toks0={#2}\toks1={#3}\calcnum*#1(#2,#3)\relax
\dd*=.23\Lengthunit\dd*=#1\dd*\divide\dd*\N*
\divide\dd*\*ths \multiply\dd*\magnitude
\d**=#1\Lengthunit\ifdim\d**<0pt\d**-\d**\fi
\multiply\N*4\divide\d**\N*\sm*\n*\*one\loop
\shl**{\dd*}\dt*=1.3\dd*\advance\n*\*one
\shl**{\dt*}\advance\n*\*one
\shl**{\dd*}\advance\n*\*two
\dd*-\dd*\ifnum\n*<\N*\repeat\n*\N*\shl**{0pt}}}

\def\w*lin(#1,#2){\rlap{\toks0={#1}\toks1={#2}\d**=\Lengthunit\dd*=-.12\d**
\divide\dd*\*ths \multiply\dd*\magnitude
\N*8\divide\d**\N*\sm*\n*\*one\loop
\shl**{\dd*}\dt*=1.3\dd*\advance\n*\*one
\shl**{\dt*}\advance\n*\*one
\shl**{\dd*}\advance\n*\*one
\shl**{0pt}\dd*-\dd*\advance\n*1\ifnum\n*<\N*\repeat}}

\def\l*arc(#1,#2)[#3][#4]{\rlap{\toks0={#1}\toks1={#2}\d**=\Lengthunit
\dd*=#3.037\d**\dd*=#4\dd*\dt*=#3.049\d**\dt*=#4\dt*\ifdim\d**>10mm\relax
\d**=.25\d**\n*\*one\shl**{-\dd*}\n*\*two\shl**{-\dt*}\n*3\relax
\shl**{-\dd*}\n*4\relax\shl**{0pt}\else
\ifdim\d**>5mm\d**=.5\d**\n*\*one\shl**{-\dt*}\n*\*two
\shl**{0pt}\else\n*\*one\shl**{0pt}\fi\fi}}

\def\d*arc(#1,#2)[#3][#4]{\rlap{\toks0={#1}\toks1={#2}\d**=\Lengthunit
\dd*=#3.037\d**\dd*=#4\dd*\d**=.25\d**\sm*\n*\*one\shl**{-\dd*}\relax
\n*3\relax\sh*(#1,#2){\xL*=\xscale\dd*\yL*=\yscale\dd*
\kern#2\xL*\lower#1\yL*\hbox{\sm*}}\n*4\relax\shl**{0pt}}}

\def\shl**#1{\c*=\the\n*\d**\d*=#1\relax
\a*=\the\toks0\c*\b*=\the\toks1\d*\advance\a*-\b*
\b*=\the\toks1\c*\d*=\the\toks0\d*\advance\b*\d*
\a*=\xscale\a*\b*=\yscale\b*
\rx* \the\cos*\a* \tmp* \the\sin*\b* \advance\rx*-\tmp*
\ry* \the\cos*\b* \tmp* \the\sin*\a* \advance\ry*\tmp*
\raise\ry*\rlap{\kern\rx*\unhcopy\spl*}}

\def\wlin*#1(#2,#3)[#4]{\rlap{\toks0={#2}\toks1={#3}\relax
\c*=#1\l*\c*\c*=.01\Lengthunit\m*\c*\divide\l*\m*
\c*=\the\Nhalfperiods5sp\multiply\c*\l*\N*\c*\divide\N*\*ths
\divide\N*2\multiply\N*2\advance\N*\*one
\dd*=.002\Lengthunit\dd*=#4\dd*\multiply\dd*\l*\divide\dd*\N*
\divide\dd*\*ths \multiply\dd*\magnitude
\d**=#1\multiply\N*4\divide\d**\N*\sm*\n*\*one\loop
\shl**{\dd*}\dt*=1.3\dd*\advance\n*\*one
\shl**{\dt*}\advance\n*\*one
\shl**{\dd*}\advance\n*\*two
\dd*-\dd*\ifnum\n*<\N*\repeat\n*\N*\shl**{0pt}}}

\def\wavebox#1{\setbox0\hbox{#1}\relax
\a*=\wd0\advance\a*14pt\b*=\ht0\advance\b*\dp0\advance\b*14pt\relax
\hbox{\kern9pt\relax
\rmov*(0pt,\ht0){\rmov*(-7pt,7pt){\wlin*\a*(1,0)[+]\wlin*\b*(0,-1)[-]}}\relax
\rmov*(\wd0,-\dp0){\rmov*(7pt,-7pt){\wlin*\a*(-1,0)[+]\wlin*\b*(0,1)[-]}}\relax
\box0\kern9pt}}

\def\rectangle#1(#2,#3){\relax
\lin#1(#2,0)\lin#1(0,#3)\mov#1(0,#3){\lin#1(#2,0)}\mov#1(#2,0){\lin#1(0,#3)}}

\def\dashrectangle#1(#2,#3){\dashlin#1(#2,0)\dashlin#1(0,#3)\relax
\mov#1(0,#3){\dashlin#1(#2,0)}\mov#1(#2,0){\dashlin#1(0,#3)}}

\def\waverectangle#1(#2,#3){\L*=#1\Lengthunit\a*=#2\L*\b*=#3\L*
\ifdim\a*<0pt\a*-\a*\def\x*{-1}\else\def\x*{1}\fi
\ifdim\b*<0pt\b*-\b*\def\y*{-1}\else\def\y*{1}\fi
\wlin*\a*(\x*,0)[-]\wlin*\b*(0,\y*)[+]\relax
\mov#1(0,#3){\wlin*\a*(\x*,0)[+]}\mov#1(#2,0){\wlin*\b*(0,\y*)[-]}}

\def\calcparab*{\ifnum\n*>\m*\k*\N*\advance\k*-\n*\else\k*\n*\fi
\a*=\the\k* sp\a*=10\a*\b*\dm*\advance\b*-\a*\k*\b*
\a*=\the\*ths\b*\divide\a*\l*\multiply\a*\k*
\divide\a*\l*\k*\*ths\r*\a*\advance\k*-\r*\dt*=\the\k*\L*}

\def\arcto#1(#2,#3)[#4]{\rlap{\toks0={#2}\toks1={#3}\calcnum*#1(#2,#3)\relax
\dm*=135sp\dm*=#1\dm*\d**=#1\Lengthunit\ifdim\dm*<0pt\dm*-\dm*\fi
\multiply\dm*\r*\a*=.3\dm*\a*=#4\a*\ifdim\a*<0pt\a*-\a*\fi
\advance\dm*\a*\N*\dm*\divide\N*10000\relax
\divide\N*2\multiply\N*2\advance\N*\*one
\L*=-.25\d**\L*=#4\L*\divide\d**\N*\divide\L*\*ths
\m*\N*\divide\m*2\dm*=\the\m*5sp\l*\dm*\sm*\n*\*one\loop
\calcparab*\shl**{-\dt*}\advance\n*1\ifnum\n*<\N*\repeat}}

\def\arrarcto#1(#2,#3)[#4]{\L*=#1\Lengthunit\L*=.54\L*
\arcto#1(#2,#3)[#4]\rmov*(#2\L*,#3\L*){\d*=.457\L*\d*=#4\d*\d**-\d*
\rmov*(#3\d**,#2\d*){\arrow.02(#2,#3)}}}

\def\dasharcto#1(#2,#3)[#4]{\rlap{\toks0={#2}\toks1={#3}\relax
\calcnum*#1(#2,#3)\dm*=\the\N*5sp\a*=.3\dm*\a*=#4\a*\ifdim\a*<0pt\a*-\a*\fi
\advance\dm*\a*\N*\dm*
\divide\N*20\multiply\N*2\advance\N*1\d**=#1\Lengthunit
\L*=-.25\d**\L*=#4\L*\divide\d**\N*\divide\L*\*ths
\m*\N*\divide\m*2\dm*=\the\m*5sp\l*\dm*
\sm*\n*\*one\loop\calcparab*
\shl**{-\dt*}\advance\n*1\ifnum\n*>\N*\else\calcparab*
\sh*(#2,#3){\xL*=#3\dt* \yL*=#2\dt*
\rx* \the\cos*\xL* \tmp* \the\sin*\yL* \advance\rx*\tmp*
\ry* \the\cos*\yL* \tmp* \the\sin*\xL* \advance\ry*-\tmp*
\kern\rx*\lower\ry*\hbox{\sm*}}\fi
\advance\n*1\ifnum\n*<\N*\repeat}}

\def\*shl*#1{\c*=\the\n*\d**\advance\c*#1\a**\d*\dt*\advance\d*#1\b**
\a*=\the\toks0\c*\b*=\the\toks1\d*\advance\a*-\b*
\b*=\the\toks1\c*\d*=\the\toks0\d*\advance\b*\d*
\rx* \the\cos*\a* \tmp* \the\sin*\b* \advance\rx*-\tmp*
\ry* \the\cos*\b* \tmp* \the\sin*\a* \advance\ry*\tmp*
\raise\ry*\rlap{\kern\rx*\unhcopy\spl*}}

\def\calcnormal*#1{\b**=10000sp\a**\b**\k*\n*\advance\k*-\m*
\multiply\a**\k*\divide\a**\m*\a**=#1\a**\ifdim\a**<0pt\a**-\a**\fi
\ifdim\a**>\b**\d*=.96\a**\advance\d*.4\b**
\else\d*=.96\b**\advance\d*.4\a**\fi
\d*=.01\d*\r*\d*\divide\a**\r*\divide\b**\r*
\ifnum\k*<0\a**-\a**\fi\d*=#1\d*\ifdim\d*<0pt\b**-\b**\fi
\k*\a**\a**=\the\k*\dd*\k*\b**\b**=\the\k*\dd*}

\def\wavearcto#1(#2,#3)[#4]{\rlap{\toks0={#2}\toks1={#3}\relax
\calcnum*#1(#2,#3)\c*=\the\N*5sp\a*=.4\c*\a*=#4\a*\ifdim\a*<0pt\a*-\a*\fi
\advance\c*\a*\N*\c*\divide\N*20\multiply\N*2\advance\N*-1\multiply\N*4\relax
\d**=#1\Lengthunit\dd*=.012\d**
\divide\dd*\*ths \multiply\dd*\magnitude
\ifdim\d**<0pt\d**-\d**\fi\L*=.25\d**
\divide\d**\N*\divide\dd*\N*\L*=#4\L*\divide\L*\*ths
\m*\N*\divide\m*2\dm*=\the\m*0sp\l*\dm*
\sm*\n*\*one\loop\calcnormal*{#4}\calcparab*
\*shl*{1}\advance\n*\*one\calcparab*
\*shl*{1.3}\advance\n*\*one\calcparab*
\*shl*{1}\advance\n*2\dd*-\dd*\ifnum\n*<\N*\repeat\n*\N*\shl**{0pt}}}

\def\triangarcto#1(#2,#3)[#4]{\rlap{\toks0={#2}\toks1={#3}\relax
\calcnum*#1(#2,#3)\c*=\the\N*5sp\a*=.4\c*\a*=#4\a*\ifdim\a*<0pt\a*-\a*\fi
\advance\c*\a*\N*\c*\divide\N*20\multiply\N*2\advance\N*-1\multiply\N*2\relax
\d**=#1\Lengthunit\dd*=.012\d**
\divide\dd*\*ths \multiply\dd*\magnitude
\ifdim\d**<0pt\d**-\d**\fi\L*=.25\d**
\divide\d**\N*\divide\dd*\N*\L*=#4\L*\divide\L*\*ths
\m*\N*\divide\m*2\dm*=\the\m*0sp\l*\dm*
\sm*\n*\*one\loop\calcnormal*{#4}\calcparab*
\*shl*{1}\advance\n*2\dd*-\dd*\ifnum\n*<\N*\repeat\n*\N*\shl**{0pt}}}

\def\hr*#1{\L*=\xscale\Lengthunit\ifnum
\angle**=0\clap{\vrule width#1\L* height.1pt}\else
\L*=#1\L*\L*=.5\L*\rmov*(-\L*,0pt){\sm*}\rmov*(\L*,0pt){\sl*}\fi}

\def\shade#1[#2]{\rlap{\Lengthunit=#1\Lengthunit
\special{em:linewidth .001pt}\relax
\mov(0,#2.05){\hr*{.994}}\mov(0,#2.1){\hr*{.980}}\relax
\mov(0,#2.15){\hr*{.953}}\mov(0,#2.2){\hr*{.916}}\relax
\mov(0,#2.25){\hr*{.867}}\mov(0,#2.3){\hr*{.798}}\relax
\mov(0,#2.35){\hr*{.715}}\mov(0,#2.4){\hr*{.603}}\relax
\mov(0,#2.45){\hr*{.435}}\special{em:linewidth \the\linwid*}}}

\def\dshade#1[#2]{\rlap{\special{em:linewidth .001pt}\relax
\Lengthunit=#1\Lengthunit\if#2-\def\t*{+}\else\def\t*{-}\fi
\mov(0,\t*.025){\relax
\mov(0,#2.05){\hr*{.995}}\mov(0,#2.1){\hr*{.988}}\relax
\mov(0,#2.15){\hr*{.969}}\mov(0,#2.2){\hr*{.937}}\relax
\mov(0,#2.25){\hr*{.893}}\mov(0,#2.3){\hr*{.836}}\relax
\mov(0,#2.35){\hr*{.760}}\mov(0,#2.4){\hr*{.662}}\relax
\mov(0,#2.45){\hr*{.531}}\mov(0,#2.5){\hr*{.320}}\relax
\special{em:linewidth \the\linwid*}}}}

\def\vdot{\rlap{\kern-1.9pt\lower1.8pt\hbox{$\scriptstyle\bullet$}}}
\def\vtimes{\rlap{\kern-3pt\lower1.8pt\hbox{$\scriptstyle\times$}}}
\def\vDot{\rlap{\kern-2.3pt\lower2.7pt\hbox{$\bullet$}}}
\def\vTimes{\rlap{\kern-3.6pt\lower2.4pt\hbox{$\times$}}}

\def\arc(#1)[#2,#3]{{\k*=#2\l*=#3\m*=\l*
\advance\m*-6\ifnum\k*>\l*\relax\else
{\rotate(#2)\mov(#1,0){\sm*}}\loop
\ifnum\k*<\m*\advance\k*5{\rotate(\k*)\mov(#1,0){\sl*}}\repeat
{\rotate(#3)\mov(#1,0){\sl*}}\fi}}

\def\dasharc(#1)[#2,#3]{{\k**=#2\n*=#3\advance\n*-1\advance\n*-\k**
\L*=1000sp\L*#1\L* \multiply\L*\n* \multiply\L*\Nhalfperiods
\divide\L*57\N*\L* \divide\N*2000\ifnum\N*=0\N*1\fi
\r*\n*  \divide\r*\N* \ifnum\r*<2\r*2\fi
\m**\r* \divide\m**2 \l**\r* \advance\l**-\m** \N*\n* \divide\N*\r*
\k**\r* \multiply\k**\N* \dn*\n*
\advance\dn*-\k** \divide\dn*2\advance\dn*\*one
\r*\l** \divide\r*2\advance\dn*\r* \advance\N*-2\k**#2\relax
\ifnum\l**<6{\rotate(#2)\mov(#1,0){\sm*}}\advance\k**\dn*
{\rotate(\k**)\mov(#1,0){\sl*}}\advance\k**\m**
{\rotate(\k**)\mov(#1,0){\sm*}}\loop
\advance\k**\l**{\rotate(\k**)\mov(#1,0){\sl*}}\advance\k**\m**
{\rotate(\k**)\mov(#1,0){\sm*}}\advance\N*-1\ifnum\N*>0\repeat
{\rotate(#3)\mov(#1,0){\sl*}}\else\advance\k**\dn*
\arc(#1)[#2,\k**]\loop\advance\k**\m** \r*\k**
\advance\k**\l** {\arc(#1)[\r*,\k**]}\relax
\advance\N*-1\ifnum\N*>0\repeat
\advance\k**\m**\arc(#1)[\k**,#3]\fi}}

\def\triangarc#1(#2)[#3,#4]{{\k**=#3\n*=#4\advance\n*-\k**
\L*=1000sp\L*#2\L* \multiply\L*\n* \multiply\L*\Nhalfperiods
\divide\L*57\N*\L* \divide\N*1000\ifnum\N*=0\N*1\fi
\d**=#2\Lengthunit \d*\d** \divide\d*57\multiply\d*\n*
\r*\n*  \divide\r*\N* \ifnum\r*<2\r*2\fi
\m**\r* \divide\m**2 \l**\r* \advance\l**-\m** \N*\n* \divide\N*\r*
\dt*\d* \divide\dt*\N* \dt*.5\dt* \dt*#1\dt*
\divide\dt*1000\multiply\dt*\magnitude
\k**\r* \multiply\k**\N* \dn*\n* \advance\dn*-\k** \divide\dn*2\relax
\r*\l** \divide\r*2\advance\dn*\r* \advance\N*-1\k**#3\relax
{\rotate(#3)\mov(#2,0){\sm*}}\advance\k**\dn*
{\rotate(\k**)\mov(#2,0){\sl*}}\advance\k**-\m**\advance\l**\m**\loop\dt*-\dt*
\d*\d** \advance\d*\dt*
\advance\k**\l**{\rotate(\k**)\rmov*(\d*,0pt){\sl*}}%
\advance\N*-1\ifnum\N*>0\repeat\advance\k**\m**
{\rotate(\k**)\mov(#2,0){\sl*}}{\rotate(#4)\mov(#2,0){\sl*}}}}

\def\wavearc#1(#2)[#3,#4]{{\k**=#3\n*=#4\advance\n*-\k**
\L*=4000sp\L*#2\L* \multiply\L*\n* \multiply\L*\Nhalfperiods
\divide\L*57\N*\L* \divide\N*1000\ifnum\N*=0\N*1\fi
\d**=#2\Lengthunit \d*\d** \divide\d*57\multiply\d*\n*
\r*\n*  \divide\r*\N* \ifnum\r*=0\r*1\fi
\m**\r* \divide\m**2 \l**\r* \advance\l**-\m** \N*\n* \divide\N*\r*
\dt*\d* \divide\dt*\N* \dt*.7\dt* \dt*#1\dt*
\divide\dt*1000\multiply\dt*\magnitude
\k**\r* \multiply\k**\N* \dn*\n* \advance\dn*-\k** \divide\dn*2\relax
\divide\N*4\advance\N*-1\k**#3\relax
{\rotate(#3)\mov(#2,0){\sm*}}\advance\k**\dn*
{\rotate(\k**)\mov(#2,0){\sl*}}\advance\k**-\m**\advance\l**\m**\loop\dt*-\dt*
\d*\d** \advance\d*\dt* \dd*\d** \advance\dd*1.3\dt*
\advance\k**\r*{\rotate(\k**)\rmov*(\d*,0pt){\sl*}}\relax
\advance\k**\r*{\rotate(\k**)\rmov*(\dd*,0pt){\sl*}}\relax
\advance\k**\r*{\rotate(\k**)\rmov*(\d*,0pt){\sl*}}\relax
\advance\k**\r*
\advance\N*-1\ifnum\N*>0\repeat\advance\k**\m**
{\rotate(\k**)\mov(#2,0){\sl*}}{\rotate(#4)\mov(#2,0){\sl*}}}}

\def\gmov*#1(#2,#3)#4{\rlap{\L*=#1\Lengthunit
\xL*=#2\L* \yL*=#3\L*
\rx* \gcos*\xL* \tmp* \gsin*\yL* \advance\rx*-\tmp*
\ry* \gcos*\yL* \tmp* \gsin*\xL* \advance\ry*\tmp*
\rx*=\xscale\rx* \ry*=\yscale\ry*
\xL* \the\cos*\rx* \tmp* \the\sin*\ry* \advance\xL*-\tmp*
\yL* \the\cos*\ry* \tmp* \the\sin*\rx* \advance\yL*\tmp*
\kern\xL*\raise\yL*\hbox{#4}}}

\def\rgmov*(#1,#2)#3{\rlap{\xL*#1\yL*#2\relax
\rx* \gcos*\xL* \tmp* \gsin*\yL* \advance\rx*-\tmp*
\ry* \gcos*\yL* \tmp* \gsin*\xL* \advance\ry*\tmp*
\rx*=\xscale\rx* \ry*=\yscale\ry*
\xL* \the\cos*\rx* \tmp* \the\sin*\ry* \advance\xL*-\tmp*
\yL* \the\cos*\ry* \tmp* \the\sin*\rx* \advance\yL*\tmp*
\kern\xL*\raise\yL*\hbox{#3}}}

\def\Earc(#1)[#2,#3][#4,#5]{{\k*=#2\l*=#3\m*=\l*
\advance\m*-6\ifnum\k*>\l*\relax\else\def\xscale{#4}\def\yscale{#5}\relax
{\angle**0\rotate(#2)}\gmov*(#1,0){\sm*}\loop
\ifnum\k*<\m*\advance\k*5\relax
{\angle**0\rotate(\k*)}\gmov*(#1,0){\sl*}\repeat
{\angle**0\rotate(#3)}\gmov*(#1,0){\sl*}\relax
\def\xscale{1}\def\yscale{1}\fi}}

\def\dashEarc(#1)[#2,#3][#4,#5]{{\k**=#2\n*=#3\advance\n*-1\advance\n*-\k**
\L*=1000sp\L*#1\L* \multiply\L*\n* \multiply\L*\Nhalfperiods
\divide\L*57\N*\L* \divide\N*2000\ifnum\N*=0\N*1\fi
\r*\n*  \divide\r*\N* \ifnum\r*<2\r*2\fi
\m**\r* \divide\m**2 \l**\r* \advance\l**-\m** \N*\n* \divide\N*\r*
\k**\r*\multiply\k**\N* \dn*\n* \advance\dn*-\k** \divide\dn*2\advance\dn*\*one
\r*\l** \divide\r*2\advance\dn*\r* \advance\N*-2\k**#2\relax
\ifnum\l**<6\def\xscale{#4}\def\yscale{#5}\relax
{\angle**0\rotate(#2)}\gmov*(#1,0){\sm*}\advance\k**\dn*
{\angle**0\rotate(\k**)}\gmov*(#1,0){\sl*}\advance\k**\m**
{\angle**0\rotate(\k**)}\gmov*(#1,0){\sm*}\loop
\advance\k**\l**{\angle**0\rotate(\k**)}\gmov*(#1,0){\sl*}\advance\k**\m**
{\angle**0\rotate(\k**)}\gmov*(#1,0){\sm*}\advance\N*-1\ifnum\N*>0\repeat
{\angle**0\rotate(#3)}\gmov*(#1,0){\sl*}\def\xscale{1}\def\yscale{1}\else
\advance\k**\dn* \Earc(#1)[#2,\k**][#4,#5]\loop\advance\k**\m** \r*\k**
\advance\k**\l** {\Earc(#1)[\r*,\k**][#4,#5]}\relax
\advance\N*-1\ifnum\N*>0\repeat
\advance\k**\m**\Earc(#1)[\k**,#3][#4,#5]\fi}}

\def\triangEarc#1(#2)[#3,#4][#5,#6]{{\k**=#3\n*=#4\advance\n*-\k**
\L*=1000sp\L*#2\L* \multiply\L*\n* \multiply\L*\Nhalfperiods
\divide\L*57\N*\L* \divide\N*1000\ifnum\N*=0\N*1\fi
\d**=#2\Lengthunit \d*\d** \divide\d*57\multiply\d*\n*
\r*\n*  \divide\r*\N* \ifnum\r*<2\r*2\fi
\m**\r* \divide\m**2 \l**\r* \advance\l**-\m** \N*\n* \divide\N*\r*
\dt*\d* \divide\dt*\N* \dt*.5\dt* \dt*#1\dt*
\divide\dt*1000\multiply\dt*\magnitude
\k**\r* \multiply\k**\N* \dn*\n* \advance\dn*-\k** \divide\dn*2\relax
\r*\l** \divide\r*2\advance\dn*\r* \advance\N*-1\k**#3\relax
\def\xscale{#5}\def\yscale{#6}\relax
{\angle**0\rotate(#3)}\gmov*(#2,0){\sm*}\advance\k**\dn*
{\angle**0\rotate(\k**)}\gmov*(#2,0){\sl*}\advance\k**-\m**
\advance\l**\m**\loop\dt*-\dt* \d*\d** \advance\d*\dt*
\advance\k**\l**{\angle**0\rotate(\k**)}\rgmov*(\d*,0pt){\sl*}\relax
\advance\N*-1\ifnum\N*>0\repeat\advance\k**\m**
{\angle**0\rotate(\k**)}\gmov*(#2,0){\sl*}\relax
{\angle**0\rotate(#4)}\gmov*(#2,0){\sl*}\def\xscale{1}\def\yscale{1}}}

\def\waveEarc#1(#2)[#3,#4][#5,#6]{{\k**=#3\n*=#4\advance\n*-\k**
\L*=4000sp\L*#2\L* \multiply\L*\n* \multiply\L*\Nhalfperiods
\divide\L*57\N*\L* \divide\N*1000\ifnum\N*=0\N*1\fi
\d**=#2\Lengthunit \d*\d** \divide\d*57\multiply\d*\n*
\r*\n*  \divide\r*\N* \ifnum\r*=0\r*1\fi
\m**\r* \divide\m**2 \l**\r* \advance\l**-\m** \N*\n* \divide\N*\r*
\dt*\d* \divide\dt*\N* \dt*.7\dt* \dt*#1\dt*
\divide\dt*1000\multiply\dt*\magnitude
\k**\r* \multiply\k**\N* \dn*\n* \advance\dn*-\k** \divide\dn*2\relax
\divide\N*4\advance\N*-1\k**#3\def\xscale{#5}\def\yscale{#6}\relax
{\angle**0\rotate(#3)}\gmov*(#2,0){\sm*}\advance\k**\dn*
{\angle**0\rotate(\k**)}\gmov*(#2,0){\sl*}\advance\k**-\m**
\advance\l**\m**\loop\dt*-\dt*
\d*\d** \advance\d*\dt* \dd*\d** \advance\dd*1.3\dt*
\advance\k**\r*{\angle**0\rotate(\k**)}\rgmov*(\d*,0pt){\sl*}\relax
\advance\k**\r*{\angle**0\rotate(\k**)}\rgmov*(\dd*,0pt){\sl*}\relax
\advance\k**\r*{\angle**0\rotate(\k**)}\rgmov*(\d*,0pt){\sl*}\relax
\advance\k**\r*
\advance\N*-1\ifnum\N*>0\repeat\advance\k**\m**
{\angle**0\rotate(\k**)}\gmov*(#2,0){\sl*}\relax
{\angle**0\rotate(#4)}\gmov*(#2,0){\sl*}\def\xscale{1}\def\yscale{1}}}

\newcount\CatcodeOfAtSign
\CatcodeOfAtSign=\the\catcode`\@
\catcode`\@=11
\def\@arc#1[#2][#3]{\rlap{\Lengthunit=#1\Lengthunit
\sm*\l*arc(#2.1914,#3.0381)[#2][#3]\relax
\mov(#2.1914,#3.0381){\l*arc(#2.1622,#3.1084)[#2][#3]}\relax
\mov(#2.3536,#3.1465){\l*arc(#2.1084,#3.1622)[#2][#3]}\relax
\mov(#2.4619,#3.3086){\l*arc(#2.0381,#3.1914)[#2][#3]}}}

\def\dash@arc#1[#2][#3]{\rlap{\Lengthunit=#1\Lengthunit
\d*arc(#2.1914,#3.0381)[#2][#3]\relax
\mov(#2.1914,#3.0381){\d*arc(#2.1622,#3.1084)[#2][#3]}\relax
\mov(#2.3536,#3.1465){\d*arc(#2.1084,#3.1622)[#2][#3]}\relax
\mov(#2.4619,#3.3086){\d*arc(#2.0381,#3.1914)[#2][#3]}}}

\def\wave@arc#1[#2][#3]{\rlap{\Lengthunit=#1\Lengthunit
\w*lin(#2.1914,#3.0381)\relax
\mov(#2.1914,#3.0381){\w*lin(#2.1622,#3.1084)}\relax
\mov(#2.3536,#3.1465){\w*lin(#2.1084,#3.1622)}\relax
\mov(#2.4619,#3.3086){\w*lin(#2.0381,#3.1914)}}}

\def\bezier#1(#2,#3)(#4,#5)(#6,#7){\N*#1\l*\N* \advance\l*\*one
\d* #4\Lengthunit \advance\d* -#2\Lengthunit \multiply\d* \*two
\b* #6\Lengthunit \advance\b* -#2\Lengthunit
\advance\b*-\d* \divide\b*\N*
\d** #5\Lengthunit \advance\d** -#3\Lengthunit \multiply\d** \*two
\b** #7\Lengthunit \advance\b** -#3\Lengthunit
\advance\b** -\d** \divide\b**\N*
\mov(#2,#3){\sm*{\loop\ifnum\m*<\l*
\a*\m*\b* \advance\a*\d* \divide\a*\N* \multiply\a*\m*
\a**\m*\b** \advance\a**\d** \divide\a**\N* \multiply\a**\m*
\rmov*(\a*,\a**){\unhcopy\spl*}\advance\m*\*one\repeat}}}

\catcode`\*=12

\newcount\n@ast

\def\n@ast@#1{\n@ast0\relax\get@ast@#1\end}
\def\get@ast@#1{\ifx#1\end\let\next\relax\else
\ifx#1*\advance\n@ast1\fi\let\next\get@ast@\fi\next}

\newif\if@up \newif\if@dwn
\def\up@down@#1{\@upfalse\@dwnfalse
\if#1u\@uptrue\fi\if#1U\@uptrue\fi\if#1+\@uptrue\fi
\if#1d\@dwntrue\fi\if#1D\@dwntrue\fi\if#1-\@dwntrue\fi}

\def\halfcirc#1(#2)[#3]{{\Lengthunit=#2\Lengthunit\up@down@{#3}\relax
\if@up\mov(0,.5){\@arc[-][-]\@arc[+][-]}\fi
\if@dwn\mov(0,-.5){\@arc[-][+]\@arc[+][+]}\fi
\def\lft{\mov(0,.5){\@arc[-][-]}\mov(0,-.5){\@arc[-][+]}}\relax
\def\rght{\mov(0,.5){\@arc[+][-]}\mov(0,-.5){\@arc[+][+]}}\relax
\if#3l\lft\fi\if#3L\lft\fi\if#3r\rght\fi\if#3R\rght\fi
\n@ast@{#1}\relax
\ifnum\n@ast>0\if@up\shade[+]\fi\if@dwn\shade[-]\fi\fi
\ifnum\n@ast>1\if@up\dshade[+]\fi\if@dwn\dshade[-]\fi\fi}}

\def\halfdashcirc(#1)[#2]{{\Lengthunit=#1\Lengthunit\up@down@{#2}\relax
\if@up\mov(0,.5){\dash@arc[-][-]\dash@arc[+][-]}\fi
\if@dwn\mov(0,-.5){\dash@arc[-][+]\dash@arc[+][+]}\fi
\def\lft{\mov(0,.5){\dash@arc[-][-]}\mov(0,-.5){\dash@arc[-][+]}}\relax
\def\rght{\mov(0,.5){\dash@arc[+][-]}\mov(0,-.5){\dash@arc[+][+]}}\relax
\if#2l\lft\fi\if#2L\lft\fi\if#2r\rght\fi\if#2R\rght\fi}}

\def\halfwavecirc(#1)[#2]{{\Lengthunit=#1\Lengthunit\up@down@{#2}\relax
\if@up\mov(0,.5){\wave@arc[-][-]\wave@arc[+][-]}\fi
\if@dwn\mov(0,-.5){\wave@arc[-][+]\wave@arc[+][+]}\fi
\def\lft{\mov(0,.5){\wave@arc[-][-]}\mov(0,-.5){\wave@arc[-][+]}}\relax
\def\rght{\mov(0,.5){\wave@arc[+][-]}\mov(0,-.5){\wave@arc[+][+]}}\relax
\if#2l\lft\fi\if#2L\lft\fi\if#2r\rght\fi\if#2R\rght\fi}}

\catcode`\*=11

\def\Circle#1(#2){\halfcirc#1(#2)[u]\halfcirc#1(#2)[d]\n@ast@{#1}\relax
\ifnum\n@ast>0\L*=\xscale\Lengthunit
\ifnum\angle**=0\clap{\vrule width#2\L* height.1pt}\else
\L*=#2\L*\L*=.5\L*\special{em:linewidth .001pt}\relax
\rmov*(-\L*,0pt){\sm*}\rmov*(\L*,0pt){\sl*}\relax
\special{em:linewidth \the\linwid*}\fi\fi}

\catcode`\*=12

\def\wavecirc(#1){\halfwavecirc(#1)[u]\halfwavecirc(#1)[d]}
\def\dashcirc(#1){\halfdashcirc(#1)[u]\halfdashcirc(#1)[d]}

\def\xscale{1}

\def\yscale{1}

\def\Ellipse#1(#2)[#3,#4]{\def\xscale{#3}\def\yscale{#4}\relax
\Circle#1(#2)\def\xscale{1}\def\yscale{1}}

\def\dashEllipse(#1)[#2,#3]{\def\xscale{#2}\def\yscale{#3}\relax
\dashcirc(#1)\def\xscale{1}\def\yscale{1}}

\def\waveEllipse(#1)[#2,#3]{\def\xscale{#2}\def\yscale{#3}\relax
\wavecirc(#1)\def\xscale{1}\def\yscale{1}}

\def\halfEllipse#1(#2)[#3][#4,#5]{\def\xscale{#4}\def\yscale{#5}\relax
\halfcirc#1(#2)[#3]\def\xscale{1}\def\yscale{1}}

\def\halfdashEllipse(#1)[#2][#3,#4]{\def\xscale{#3}\def\yscale{#4}\relax
\halfdashcirc(#1)[#2]\def\xscale{1}\def\yscale{1}}

\def\halfwaveEllipse(#1)[#2][#3,#4]{\def\xscale{#3}\def\yscale{#4}\relax
\halfwavecirc(#1)[#2]\def\xscale{1}\def\yscale{1}}

\catcode`\@=\the\CatcodeOfAtSign

\title{Lorentz breaking supersymmetry and Horava-Lifshitz-like models}

\author{M. Gomes}
\affiliation{Instituto de F\'\i sica, Universidade de S\~ao Paulo\\
Caixa Postal 66318, 05315-970, S\~ao Paulo, SP, Brazil}
\email{mgomes, queiruga, ajsilva@if.usp.br}

\author{J. Queiruga}

\affiliation{Instituto de F\'\i sica, Universidade de S\~ao Paulo\\
Caixa Postal 66318, 05315-970, S\~ao Paulo, SP, Brazil}
\email{queiruga@if.usp.br}

\author{A. J. da Silva}
\affiliation{Instituto de F\'\i sica, Universidade de S\~ao Paulo\\
Caixa Postal 66318, 05315-970, S\~ao Paulo, SP, Brazil}
\email{ajsilva@if.usp.br}

\begin{abstract}
We present a Lorentz-breaking supersymmetric algebra characterized by a critical exponent $z$. Such construction requires a nontrivial modification of the supercharges and superderivatives. The improvement of renormalizability for supersymmetric scalar QED is shown and the K\"ahlerian effective potentials are calculated in different cases. We also show how the theory flows naturally to the Lorentz symmetric case at low energies.
\end{abstract}

\maketitle

\section{Introduction}

It is known that the addition of higher-derivative corrections  improves the ultraviolet (UV) behavior of generic quantum field theories
allowing in particular the construction of a renormalizable class of quantum gravity models \cite{Stelle}. However,
 usually the price  to pay  is the appearance of ghost excitations and the breaking of unitarity \cite{Hawking,antoniadis}. One possible way out of this problem consists of assuming scaling properties which are anisotropic between space and time, so that only high spatial derivatives are introduced, the so-called Horava-Lifshitz (HL) models (see \cite{Lifshitz,anselmi,Horava2,Horava}). Such anisotropy is characterized by a critical exponent $z$, related with the degree of the highest spatial derivative in such way that 
 for  $z=1$, the isotropy between time and space holds and the theory is Lorentz-invariant. For other values of $z$, Lorentz symmetry is broken although it is expected  to be restored at low energies \cite{Iengo,Pedro}. This procedure has been applied to models without or with gauge symmetry and also to supersymmetric theories.
 
 Concerning the building of HL like supersymmetric   models, two approaches have been proposed,  accordingly
 the superalgebra has the standard form \cite{Petrov6} or was deformed to accommodate higher spatial derivatives \cite{redigolo}. In the first
 situation, modifications are made directly in the  action leading to terms containing  both time and spatial derivatives which break the simplicity of the original proposal.  On the other hand, the inclusion of high spatial derivatives in the generators  of the superalgebra has a drawback in the sense  that they do not verify Leibniz rule making difficult to introduce self-interactions of chiral (or antichiral) superfields. We will show that, in spite of this difficulty, it is possible to formulate a theory free of pathologies. The crucial observation to take into account is that, although the self-interacting terms constructed as products of at least three chiral quantities are not chiral any longer, they  do not break the supersymmetry (SUSY) if they are integrated over the whole Grassmann space. Usually, for $z=1$ theories, such procedure leads to non-renormalizable models but in the anisotropic situation $z>1$, because of the ultraviolet improvement of the propagators, they may be allowed. In this work, we study  a HL like version  of supersymmetric QED constructed along these lines,
 determine its K\"ahlerian effective  potential and discuss the emergence of Lorentz symmetry at low energies.

This paper is organized as follows. In Sec. II we present a higher spatial derivative supersymmetric algebra and some considerations related with it. In Sect. III we supersymmetrize an anisotropic version of scalar QED and determine the superpropagators and the superficial degree of divergence of the theory. Sect. IV is devoted to one-loop calculations: the K\"ahlerian effective potential, for ${\cal N}=1$  SUSY and the analysis of the restoration of Lorentz symmetry at low energies. Finally, in Sect. V we present our conclusions. In an Appendix we collected some details of the calculations.

\section{Lorentz-Violating SUSY}

We are interested in the formulation of supersymmetric theories  which behave anisotropically under a generic scaling, $x^{i}\rightarrow b x^{i}$,  $t\rightarrow b^{z}t$. We recall that
the usual Lorentz-invariant supersymmetric algebra is defined by the following anticommutation relation of the supercharges (we follow the notation of \cite{Bagger}):

\be
\lbrace Q_\alpha, \bar{Q}_{\dot{\alpha}} \rbrace=2i\sigma_{\alpha\dot{\alpha}}^\mu\pa_\mu,
\ee
\noindent
where

\bea
Q_\alpha&=&\frac{\pa}{\pa\theta^\alpha}-i\sigma_{\alpha\dot{\alpha}}^\mu\bar{\theta}^{\dot{\alpha}}\pa_\mu,\\
\bar{Q}_{\dot{\alpha}}&=&-\frac{\pa}{\pa\bar{\theta}^{\dot{\alpha}}}+i\theta^\alpha \sigma_{\alpha\dot{\alpha}}^\mu\pa_\mu.
\eea

In this case, the anticommutator $\{Q_\alpha,\bar{Q}_{\dot{\alpha}}\}$ is proportional to $P_\mu=i\pa_\mu$ but, in the Lorentz-violating case, the Coleman-Mandula theorem is compatible with a larger class of superalgebras, allowing for a more general anticommutation relation \cite{redigolo}:

\be
\lbrace Q_\alpha, \bar{Q}_{\dot{\alpha}} \rbrace=2\sigma_{\alpha\dot{\alpha}}^{ \mu'} P_\mu+2\eta \sigma ^i_{\alpha\dot{\alpha}}\mathcal{O}_i,\label{algm}
\ee
\noindent
where $\sigma^{0'}=\sigma^0$ and $\sigma^{i'}=c\sigma^i$, the dimension of the constant $c$ is $\lbrack c \rbrack =z-1 $ and $\eta$ is a free parameter measuring the breaking of Lorentz invariance. The operator $\mathcal{O}_i$ will depend in general on the spatial part of the momentum (Lorentz violating SUSY in isotropic space was treated in \cite{Berger1}). For our purposes we are interested in the following operator 

\be
\mathcal{O}_i=\Delta^{\frac{z-1}{2}}\partial_i,
\ee

\noindent
where $\Delta=-\pa_i \pa_i$. The Lorentz-breaking parameter is $z$, and when $z\rightarrow 1$ we recover the usual Lorentz-invariant algebra up to an appropriate redefinition of the constants . This new term in the superalgebra requires a modification of the supercharges (for more details see \cite{redigolo}):

\bea
Q_\alpha&=&\frac{\pa}{\pa\theta^\alpha}-i\sigma_{\alpha\dot{\alpha}}^{\mu '}\bar{\theta}^{\dot{\alpha}}\partial_\mu-i\eta\sigma_{\alpha\dot{\alpha}}^i\bar{\theta}^{\dot{\alpha}}\Delta^{\frac{z-1}{2}}\partial_i,\label{cha1}\\
\bar{Q}_{\dot\alpha}&=&-\frac{\pa}{\pa\bar{\theta}^{\dot{\alpha}}}+i\theta^\alpha \sigma_{\alpha\dot{\alpha}}^{\mu '}\pa_\mu
+i\eta\sigma_{\alpha\dot{\alpha}}^i\bar{\theta}^{\dot{\alpha}}\Delta^{\frac{z-1}{2}}\partial_i.
\label{cha2}
\eea
\noindent
and superderivatives:

\bea
D_\alpha&=&\frac{\pa}{\pa\theta^\alpha}+i\sigma_{\alpha\dot{\alpha}}^{\mu'}\bar{\theta}^{\dot{\alpha}}\partial_\mu+i\eta\sigma_{\alpha\dot{\alpha}}^i\bar{\theta}^{\dot{\alpha}}\Delta^{\frac{z-1}{2}}\partial_i,\\
\bar{D}_{\dot\alpha}&=&-\frac{\pa}{\pa\bar{\theta}^{\dot{\alpha}}}-i\theta^\alpha \sigma_{\alpha\dot{\alpha}}^{\mu'}\pa_\mu
-i\eta\sigma_{\alpha\dot{\alpha}}^i\bar{\theta}^{\dot{\alpha}}\Delta^{\frac{z-1}{2}}\partial_i.
\label{derm}
\eea

\noindent
Note that the $\eta$ parameter is dimensionless for this choice of $\mathcal{O}_i$ operator. Since for $z \neq 1$ the operators $Q_\alpha$, $\bar{Q}_{\dot{\alpha}}$,  and  the corresponding superderivatives do not obey the Leibniz rule, we cannot define chiral superfields in the usual way. Indeed,  the deformed supercharges (\ref{cha1}) and (\ref{cha2}) require the modification of the chiral superfields:

\be
\tilde{\Phi}=\phi+i\theta\sigma^\mu\bar{\theta}\tilde{\pa}_\mu\phi+\frac{1}{4}\theta\theta\bar{\theta}\bar{\theta}\tilde{\square}\phi+\sqrt{2}\theta\psi-\frac{i}{\sqrt{2}}\theta\theta\tilde{\pa}_\mu\psi\sigma^\mu\bar{\theta}+\theta\theta F,\label{chiral}
\ee
\noindent
where $\tilde{\partial_0}=\partial_0,\quad \tilde{\partial_i}=c\partial_i +\eta\Delta^{\frac{z-1}{2}}\partial_i\label{deriv}$ and the modified D'Alembertian  $\tilde{\square}\equiv \tilde{\partial}_{0}^{2}-\tilde{\partial}_{i}\tilde{\partial}_{i}=\partial_0^2+c^2\Delta +2c\eta \Delta^{\frac{z+1}{2}}+\eta^2\Delta^{z}$. The modified superfield satisfies the chirality condition  $\bar{D}\Phi=0$.

Besides that, we introduce a vector superfield
\be
V=-\theta \sigma^\mu\bar{\theta} \tilde{A}_\mu+i\theta\theta\bar{\theta}\bar{\lambda}-i\bar{\theta}\bar{\theta}\theta\lambda+\frac{1}{2}\theta\theta\bar{\theta}\bar{\theta}D\label{vector},
\ee
\noindent
where $\tilde{A}_0=A_0,\quad\tilde{A}_i=cA_i+\eta\Delta^{\frac{z-1}{2}}A_i \label{gauge}.$

This is  a natural modification of the gauge field preserving the gauge symmetry (a deformation only on the derivatives would lead to the  breaking of gauge symmetry). Note that  the deformed field strength $\tilde{F}_{\mu\nu}=\tilde{\partial}_\mu\tilde{A}_\nu-\tilde{\partial}_\nu\tilde{A}_\mu$ is invariant under the usual gauge transformation $A_\mu\rightarrow A_\mu+\partial_\mu\Lambda$. 

Because of the fact that $\bar{D}_{\dot{\alpha}}(\Phi_1 \Phi_2)\neq \bar{D}_{\dot{\alpha}}(\Phi_1)\Phi_2+\Phi_1\bar{D}_{\dot{\alpha}}(\Phi_2)$, it is impossible to define chiral  composed operators  as products of chiral superfields.
In this context, a question that naturally arises  concerns  the class of supersymmetric actions one may construct with this deformed algebra. Now, every D-term (obtained after integrating over all Grassmann space) is still supersymmetric. Let us analyze the F-terms. For a general supersymmetric variation $\delta=\epsilon^\alpha Q_\alpha +\bar{\epsilon}_{\dot{\alpha}}\bar{Q}^{\dot{\alpha}}$ and a general chiral F-term $\int d^2\theta W(\Phi)$, we have



\be
\delta\int d^2\theta W(\Phi)=\int d^2\theta \left(\epsilon^\alpha Q_\alpha W(\Phi)+\bar{\epsilon}_{\dot{\alpha}}\bar{Q}^{\dot{\alpha}}W(\Phi)\right).
\ee
\noindent

The term proportional to $\epsilon^\alpha$ vanishes because differentiation with respect to $\theta$ gives zero  after  the          $\theta$-integration,

\bea
\int d^2\theta \frac{\pa}{\pa \theta} \text{(Something)}&=&0,
\eea
and the term proportional to $\bar \theta$ is a total derivative vanishing after integration on  the coordinate space:
\bea
\int d^{4}x\int d^2\theta (-i\sigma_{\alpha\dot{\alpha}}^{\mu'}\bar{\theta}^{\dot{\alpha}}\partial_\mu-i\sigma_{\alpha\dot{\alpha}}^i\bar{\theta}^{\dot{\alpha}}\Delta^{\frac{z-1}{2}}\partial_i)\text{(Something)}&=&0.
\eea

Let us now analyze the term proportional to $\bar{\epsilon}_{\dot{\alpha}}$:

\bea
\delta\vert_{\bar{\epsilon}}\int d^2 \theta W(\Phi)&\equiv&\bar{\epsilon}_{\dot{\alpha}}\int d^2\theta \bar{Q}^{\dot{\alpha}} W(\Phi)=\bar{\epsilon}^{\dot{\alpha}}\int d^2\theta(-i\sigma_{\alpha\dot{\alpha}}^{\mu'}\bar{\theta}^{\dot{\alpha}}\partial_\mu-i\sigma_{\alpha\dot{\alpha}}^i\bar{\theta}^{\dot{\alpha}}\Delta^{\frac{z-1}{2}}\partial_i) W(\Phi)\nonumber\\
&-&\bar{\epsilon}^{\dot{\alpha}}\int d^2\theta \frac{\pa}{\pa\bar{\theta}^{\dot{\alpha}}}W(\Phi).
\eea

The last term in the above relation  in general is not  a total derivative, therefore:

\be
\delta\vert_{\bar{\epsilon}}\int d^2\theta W(\Phi)=\text{total derivative}-\bar{\epsilon}^{\dot{\alpha}}\int d^2\theta \frac{\pa}{\pa\bar{\theta}^{\dot{\alpha}}}W(\Phi).
\ee

For $W(\Phi)=\Phi^n$ we get schematically:
\be
\int d^2\theta \frac{\pa}{\pa\bar{\theta}^{\dot{\alpha}}}\Phi^n\propto (n-1)\phi^{n-2}\mathcal{O}_\mu\phi \psi_\alpha+\phi^{n-1}\mathcal{O}_\mu\psi_\alpha.
\ee

For $\mathcal{O}_\mu=\partial_\mu$, as in the Lorentz invariant situation, the above expression is a total derivative, and the SUSY is preserved. If $\mathcal{O}_\mu$ is a higher derivative operator (and this is our case), the above expression is not a total derivative for $n>2$. The consequence of this observation, as it was pointed out by \cite{redigolo}, is that our Lorentz violating SUSY does not allow for $F-terms$ of degree greater that $2$, hence self interacting chiral terms explicitly break the SUSY.

\subsection{DEFORMED PROJECTION OPERATORS}

Projection operators usually simplify superfield calculations in Lorentz-invariant supersymmetric theories. The ones we are interested in are the following

\be
\Pi_{1}=\frac{D^{2}\bar{D}^{2}}{16\square},\quad\Pi_{2}=\frac{\bar{D}^{2}{D}^{2}}{16\square},\quad\Pi_0=\frac{1}{16}\frac{\lbrace D^2,\bar{D}^2\rbrace}{\square},\quad \Pi_{1/2}=-\frac{1}{8\square}D^\alpha\bar{D}^2D_\alpha
\ee

As we will see below we only need the multiplication rule for the last two of them. Those operators are idempotents and their cross composition vanishes,

\be
 \Pi_i \Pi_j=\delta_{ij}\Pi_i,\quad i, j=0,1,2,1/2\label{pro}.
\ee

We want to show that when we deformed these operators (i.e. $\pa_\mu\rightarrow \tilde{\pa}_\mu$) these properties still hold. First we write the new projectors in terms of the deformed superderivatives and D'Alembertians:

\be
\tilde{\Pi}_0=\frac{1}{16}\frac{\lbrace \tilde{D}^2,\bar{\tilde{D}}^2\rbrace}{\tilde{\square}},\quad \tilde{\Pi}_{1/2}=-\frac{1}{8\tilde{\square}}\tilde{D}^\alpha\bar{\tilde{D}}^2\tilde{D}_\alpha,
\ee

\noindent
but since $\lbrack \frac{\pa}{\pa\theta^\alpha},\tilde{\partial}_\mu  \rbrack=0$ and $\lbrack \tilde{\partial}_\mu  ,\tilde{\partial}_\nu  \rbrack=0$, the relation for tilde operators is the same as for the usual ones (\ref{pro}),

\be
 \tilde{\Pi}_i \tilde{\Pi}_j=\delta_{ij}\tilde{\Pi}_i,\quad i, j=0,1,2,1/2\label{prom}.
\ee

 From now on, we will omit tildes  in the  projectors and superderivatives, all quantities being understood to be the tilde ones. With this supersymmetric structure we are ready to do some calculations.

\section{Supersymmetric HL-like electrodynamics}

The model we are going to analyze is a Lorentz broken version of  supersymmetric scalar  QED.   Without SUSY, variants of this model with different critical exponents and non local operators were treated in the literature (see for example \cite{farakos1,farakos2,farakos3,Petrov5,Petrov8}).

The Lagrangian for an Abelian gauge theory  with an arbitrary critical exponent $z$  looks like:

\bea
\mathcal{L}=\frac{1}{2}c^2 F_{0i}F^{0i}-\frac{1}{4}c^4F_{ij}F^{ij}+\frac{1}{2}F_{0i}Q_1 (\Delta;z)F^{0i}-\frac{1}{4}c^4F_{ij}Q_2(\Delta;z)F^{ij}\label{yang}.
\eea

The first two terms in the above expression correspond to the usual Abelian gauge theory with the appropriate anisotropic scaling. The other two terms constitute  higher derivative corrections. The operators $Q_1 (\Delta;z)$ and $Q_2(\Delta;z)$ are polynomials in $\Delta$, with increasing degree in $z$ ($\ge1$); we will fix them in order to obtain the supersymmetric extension.  We can add a minimally coupled complex field $\phi$, 

\be
\mathcal{L}_2=-\bar{\phi}\left( \partial_0^2-c^2\partial_i^2 \right)\phi-\bar{\phi} Q_1 (\Delta;z)\pa_i^2\phi +\text{gauge interactions}\label{quad}.
\ee

The gauge  interactions will be introduced by promoting the usual interaction term $igA_\mu (\bar{\phi}\pa^\mu\phi-\pa^\mu\bar{\phi}\phi)+g^2 A_\mu A^\mu \bar{\phi}\phi$ to the deformed version, by replacing gauge  fields and operators as defined in the text after equations (\ref{chiral}) and (\ref{vector}).
We need also to add a gauge fixing term. A natural extension of the Feynman gauge to our anisotropic space has the  form

\be
\mathcal{L}_{gf}=-\frac{1}{2}\left( \pa_0A_0-c^2\pa_i A_i-Q_1(\Delta;z)\pa_i A_i \right)^2\label{gf}.
\ee

This gauge fixing  eliminates the cross terms of the gauge fields,  without the introduction of  non local terms in the action (see for example \cite{farakos3,Petrov5}).
The free propagators of the theory are

\bea
\langle \phi\bar{\phi}   \rangle&=&\frac{i}{(k_0^2-c^2 \bar{k}^2)-Q_1(\bar{k}^2;z)\bar{k}^2} ,\\
\langle A_0 A_0  \rangle&=&\frac{-i}{(k_0^2-c^2 \bar{k}^2)-Q_1(\bar{k}^2;z)\bar{k}^2}, \\
\langle A_i A_j  \rangle&=&\frac{-i\,\delta_{ij}}{(k_0^2-c^2 \bar{k}^2)-Q_1(\bar{k}^2;z)k_0^2-Q_2(\bar{k}^2;z)\bar{k}^2}  \quad.
\eea

Thus, if we choose the polynomials $Q_i(\bar{k}^2;1)=\alpha$, $\alpha$ being a dimensionless constant, then after a redefinition of the variables to absorb the constants, the propagators have exactly the standard form. If, for a suitable choice of $z$, the degree of $Q_i(\bar{k}^2;z)$ in $\bar{k}^2$ is greater than one then, at high energy, the propagators are dominated by this anisotropic term.

Our  next step is to build the supersymmetric version of this family of models. 
The nice point of our ``deformed" SUSY is that it provides a natural formulation for this class of theories. The supersymmetric version of the model (\ref{yang})+(\ref{quad}) has the  familiar form:

\be
\mathcal{L}=\frac{1}{4}\left(\int d^2\theta W^\alpha W_\alpha+\int d^2\bar{\theta} \bar{W}^{\dot{\alpha}} \bar{W}_{\dot{\alpha}}\right)+\int d^2\theta d^2\bar{\theta}\bar{\Phi}e^{gV}\Phi,
\ee

\noindent
where all the superfields are understood to be the ones corresponding to the modified superalgebra (\ref{chiral}) and  (\ref{vector})  such that the superfield strengths are

\bea
W_{\alpha}=\bar{D}^2 D_\alpha V.
\eea

For the gauge fixing action we choose

\be
\mathcal{L}_{gf}=\frac{1}{16\lambda}\int d^2\theta d^2\bar{\theta} (\bar{D}^2 V)D^2V.
\ee

If we expand in components, the bosonic sector (with $F=D=0$) of this theory corresponds to (\ref{yang}) and (\ref{quad}) where:

\be
Q_1(\Delta;z)=\mathcal{P}(\Delta;z)^2-c^2, \quad Q_2(\Delta;z)=\mathcal{P}(\Delta;z)^4-c^4\label{poly},
\ee

\noindent
being $\mathcal{P}(\Delta;z)=c+\eta\Delta^{\frac{z-1}{2}}$. Note that it is possible to change these higher derivative corrections in the supersymmetric theory by changing the superalgebra adding other spatial operators. Such a change modifies the polynomial $\mathcal{P}(\Delta;z)$ but the forms (\ref{poly}) of $Q_1$ and $Q_2$  remain unchanged.

The superpropagators are then

\bea
\langle  \Phi(\xi_1)\bar{\Phi}(\xi_2) \rangle&=&i\frac{\bar{D}^2D^2}{16\tilde{\square}}\delta^8(\xi_1-\xi_2)\label{ss1},\\
\langle V(\xi_1)V(\xi_1)\rangle&=&-\frac{i}{\tilde{\square}}\left(\lambda\frac{\lbrace D^2,\bar{D}^2\rbrace}{16\tilde{\square}}-\frac{D^\alpha \bar{D}^2D_\alpha}{8\tilde{\square}}\right)\delta^8(\xi_1-\xi_2)\label{ss2},
\eea
where $\xi=(x,\theta,\bar\theta)$.


All the observations we made for the propagators also hold for the superpropagators. It also be also useful to write (\ref{ss1}) and (\ref{ss2}) in terms of the modified projection operators (section II.A)

\bea
&\langle  \Phi(\xi_1)\bar{\Phi}(\xi_2) \rangle&=i\Pi_1\delta^8(\xi_1-\xi_2)\label{s1}\\
&\langle V(\xi_1)V(\xi_2)\rangle&=-i\frac{1}{\tilde{\square}}\left(\Pi_{1/2}+\lambda\Pi_0\right)\delta^8(\xi_1-\xi_2)\label{s2}
\eea

\subsection{SUPERFICIAL DEGREE OF DIVERGENCE}

With the explicit expressions of the superpropagators we may determine the superficial degree of divergence (SDD). Because of the  anisotropic scaling between time and space coordinates, the dimensions of the derivatives are $\lbrack\partial_0\rbrack=z$ and $\lbrack\partial_i\rbrack=1$, and therefore, in $d$ spatial dimensions we have 

\be
\lbrack A_0\rbrack=\frac{d+z}{2}-1,\quad \lbrack A_i\rbrack=\frac{d-z}{2},
\ee

and 

\be
\lbrack  \phi\rbrack =\lbrack  \bar{\phi}\rbrack=\frac{d-z}{2},\quad \lbrack g \rbrack=\frac{z-d}{2}+1.
\ee

However, the deformed operators in the supersymmetric algebra are such that the dimensions of space and time components are equal and the anisotropy is compensated (since $\lbrack c  \rbrack=z-1$):

\be
\lbrack  \tilde{\partial}_0\rbrack =\lbrack  \tilde{\partial}_i\rbrack=z,\quad  \lbrack  \tilde{A}_0\rbrack =\lbrack  \tilde{A}_i\rbrack=\frac{d-z}{2}.
\ee

Let us consider a supergraph with $L$ loops and $P$ internal propagators. From the
 integration over the internal momenta, each loop contributes to the SDD with a factor $(d+z)$, so that the total  contribution will be $(d+z)L$. For the power counting, it is convenient to associate the $D^{2}$ and  $\bar{D^{2}}$ factors in the scalar propagator with the vertices joined by the corresponding line. Thus, each propagator,  $\langle VV\rangle$ or $\langle \Phi\bar{\Phi}  \rangle$ contributes with a factor $-2z$, and to the SDD with $-2zP$. From each vertex, and due to the factors $D^2$ and $\bar{D}^2$, we get an extra contribution $2zV$, unless the vertex is connected to a external chiral or antichiral line, which gives an extra factor $-zE_c$. Besides that,  taking into account the following fundamental relation
 
\be
\delta_{12}D^2\bar{D}^2\delta_{12}=16\delta_{12},
\ee
\noindent
we have to subtract  $2z$ for every loop, and thus  $-2zL$ for the SDD. Putting all together, the SDD is given by

\be
\omega=(d+z)L-2zP-2zL+2zV-zE_c.
\ee

Finally, the use of  the topological relation $L+V-P=1$ yields

\be
\omega=(d-z)+(d-3z)(P-V)-zE_c,
\ee
\noindent
so that, for $d=3$, 

\be
\omega=(3-z)+3(1-z)(P-V)-zE_c.
\ee

Note that for $z=1$ we recover the SDD of the Lorentz invariant theory. The model is superrenormalizable  when $z> 1$. As expected, the behavior of the theory is improved by the critical exponent $z$. For example, in the relativistic theory ($z=1$) tadpole diagrams with one external leg diverge linearly,  but in the LV theory for $z>3/2$ they converge; also, in the relativistic context graphs with two external legs diverge logarithmically, but in the LV theory with $z>1$ such diagrams are convergent.

\section{ONE-LOOP CALCULATIONS AND THE K\"AHLER POTENTIAL}

\begin{figure}[h]
    \centering
    \includegraphics[width=0.6\textwidth]{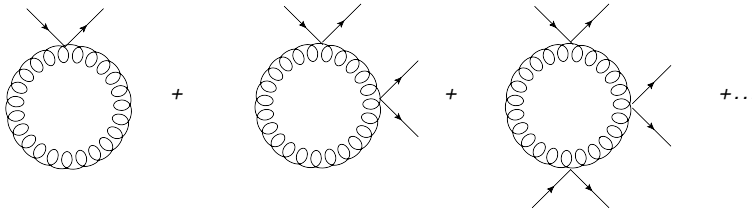}
    \caption{One-loop gauge field contribution: $\mathcal{K}_a$}
    \label{fig1}
\end{figure}

With the superpropagators written in terms of the projection operators we can determine the K\"ahler potential. For details in the computation of K\"ahler  potentials see \cite{YMEP1,YMEP2,YMEP3}. Our calculation follows closely  the one presented in \cite{Petrov7}, but with the modified SUSY in anisotropic space-time. First we need to sum over all one-loop diagrams with n-$\bar{\Phi}\Phi$ external legs and containing  only internal gauge propagators, as show in Fig. \ref{fig1},
which corresponds to the following series:

\be
i\,\mathcal{K}_a=\int d^8\xi \sum_{n=1}^\infty\frac{(-1)^n}{2n}\left(  \frac{g^2\bar{\Phi}\Phi}{\tilde{\square}}\left(\Pi_{1/2}+\lambda\Pi_0 \right)  \right)^n\delta^8(\xi_1-\xi_2) .
\ee

\begin{figure}[h]
    \centering
    \includegraphics[width=0.5\textwidth]{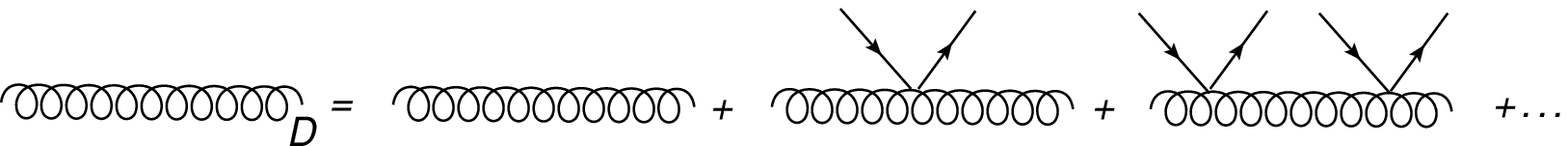}
    \caption{Dressed gauge propagator}
    \label{fig2}
\end{figure}

By carrying out the Fourier transformation, performing a  Wick rotation and taking into account the algebra (\ref{prom}) of the projection operators which satisfy the relations $\tilde{\square}\Pi_0\delta^8(\xi_1-\xi_2)\vert_{\theta_1=\theta_2}=2$ and $\tilde{\square}\Pi_{1/2}\delta^8(\xi_1-\xi_2)\vert_{\theta_1=\theta_2}=-2$, we get

\be
\mathcal{K}_a=\int d^8\xi \int \frac{d^4 k_E}{(2 \pi)^4}\frac{1}{\tilde{k}_E^2}\left(\ln (1+\frac{g^2\bar{\Phi}\Phi}{\tilde{k}_E^2})-\ln (1+\frac{\lambda g^2\bar{\Phi}\Phi}{\tilde{k}_E^2})  \right), \label{ka}
\ee
\noindent
where $\tilde{k}_E^2=k_0^2+c^2\textbf{k}^2+2c\eta(\textbf{k}^2)^{\frac{z+1}{2}}+\eta^2(\textbf{k}^2)^z$. At the one-loop level, there is yet another family of diagrams contributing to the K\"ahler  potential involving both $V$ and $\Phi$ propagators,
but, before considering that, we need to add all possible insertions of $\bar{\Phi}\Phi$ external legs in all $V$-propagators.
We can perform this sum by introducing the ``dressed" propagator \cite{Petrov5}, consisting of the $VV$-propagator with all possible insertions of $\bar{\Phi}\Phi$ external legs (Fig. \ref{fig2}).

\vspace{1cm}

The dressed propagator has the following expression:

\bea
\langle V(\xi_1) V(\xi_2)   \rangle_D&=&\langle V V   \rangle\sum_{n=0}^\infty\left( g^2\bar{\Phi}\Phi\langle V V   \rangle\right)^n\delta^8(\xi_1-\xi_2)=\nonumber \\
&=&-\sum_{n=0}^\infty (g^2\bar{\Phi}\Phi)^n\left(\frac{1}{\tilde{\square}}\right)^{n+1}(\Pi_{1/2}+\lambda^{n+1}\Pi_0)\delta^8(\xi_1-\xi_2)=\nonumber\\
&=&-\left( \frac{1}{\tilde{\square}+g^2\bar{\Phi}\Phi}\Pi_{1/2}+\frac{\lambda}{\tilde{\square}+\lambda g^2\bar{\Phi}\Phi}\Pi_0  \right)\delta^8 (\xi_1-\xi_2).
\eea

\begin{figure}[h]
    \centering
   \includegraphics[width=0.5\textwidth]{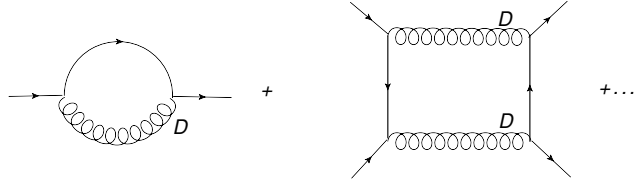}
    \caption{One-loop gauge-boson contribution: $\mathcal{K}_b$}
    \label{fig3}
\end{figure}

 Now, if we insert the dressed propagator in the one-loop diagrams  involving also $\Phi$-propagators, we have
the series shown in Fig. \ref{fig3}, which gives

\be
i\,\mathcal{K}_b=\int d^8 \xi_1 \sum_{n=1}^\infty\frac{1}{2n}\left( g^2 \bar{\Phi}\Phi\Pi_0\langle V V \rangle_D \right)^n\delta^8(\xi_1-\xi_2)\vert_{\theta_1=\theta_2}, \label{kb}
\ee

\noindent
where $\langle V V \rangle_D$ is the dressed propagator excluding the $\delta$ factor. By  using  properties of the projection operators, after performing a Fourier transformation and Wick rotation, we arrive at,

\be
\mathcal{K}_b=\int d^8 \xi \int \frac{d^4 k_E}{(2\pi)^4}\frac{1}{\tilde{k}_E^2} \ln\left(1+\frac{\lambda g^2\bar{\Phi}{\Phi}}{\tilde{k}_E^2+\lambda g^2\bar{\Phi}\Phi}\right) .
\ee

The total K\"ahler  potential is the sum of both contributions, $\mathcal{K}^{one-loop}(z,\lambda)=\mathcal{K}_a^{one-loop}+\mathcal{K}_b^{one-loop}\label{totalpot}$. For the sake of simplicity, we will work in the Landau gauge, $\lambda=0$,
where the potential results in:
\be
\mathcal{K}=\int d^8 \xi\int \frac{d^4 k_E}{(2\pi)^4}\frac{1}{\tilde{k}_E^2} \ln\left(1+
\frac{\mu^2}{\tilde{k}_E^2}\right) ,\label{potK}
\ee
where we defined: $\mu^2=g^2\bar{\Phi}{\Phi}$.  Up to a $\mu$ independent constant, by carrying out the integrations on the angles and in $k_0$, we arrive at
\be
\mathcal{K}= \int d^{8}\xi\left[ \frac{1}{2\pi^{2}}\int_{0}^{\infty} dk \frac{k^{2}}{\Delta} \ln\left(\frac12+\frac12\sqrt{1+\frac{\mu^{2}}{\Delta^{2}}}\right )\right],\label{potK1}
\ee
where $\Delta= c k +\eta k^{z}$. Unless for $z=1$, the $k$ integral does not have a simple analytic expression. However, by straightforward
power counting it is easy to see that it is convergent if  $z>1$ but logarithmically divergent if $z=1$. Actually, for $z=1$, using dimensional regularization,  we get

\be
\mathcal{K}(z=1,\lambda=0)=-\frac{1}{16 \pi ^2  (c+\eta )^3}\int d^{8}\xi\left [\frac{2\, g^{2}\Phi  \bar{\Phi }}{{d-3}}+g^2  \bar{\Phi }\Phi  \log \left(\frac{g^{2}\Phi  \bar{\Phi }}{4 \pi(c+\eta)^{2}  e^{2-\gamma }}\right)\right],
\ee
where, as usual, the divergence manifests itself as a pole in $d=3$. A mass renormalization is necessary, to eliminate the pole $1/(d-3)$  and 
define the potential.

\subsection{ADDING MORE INTERACTIONS}

\begin{figure}[h]
    \centering
    \includegraphics[width=0.4\textwidth]{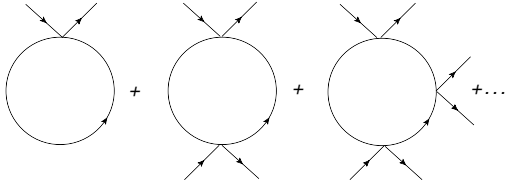}
    \caption{One-loop selfinteracting contribution}
    \label{fig5}
\end{figure}

Despite the fact that this anisotropic SUSY does not allow for chiral interactions, i.e. terms of the form $\int d^2\theta \mathcal{P}(\Phi)+h.c.$ ($\mathcal{P}$ being a polynomial of degree greater than 2), it is possible to introduce derivative interactions as D-terms by integrating over the whole Grassmann space.  We may add for example a quartic interaction of the form $\int d^2\theta d^2\bar{\theta}D^\alpha \Phi D_\alpha \Phi \bar{D}_{\dot{\beta}}\bar{\Phi} \bar{D}^{\dot{\beta}}\bar{\Phi}$ or $\int d^2\theta d^2\bar{\theta} \bar{\Phi}\Phi\bar{\Phi}\Phi$. 
These kind of terms were introduced in \cite{ovrutgal}, \cite{ovrutghost}, \cite{kehagias} and \cite{queiruga} in the context of supersymmetric galileons, SUSY ghost condensates, higher derivative supergravity and SUSY Skyrme models, respectively.  Of course, they are derivative interactions of at least fourth order, since the integration over the full Grassmann space always generates derivatives. To determine the effective K\"ahler potential for the quartic self-interaction,  we need to add to the effective K\"ahler potential  calculated before one extra family of diagrams consisting of one loop on $\Phi$ and $n$ $\bar{\Phi}\Phi$ external legs, corresponding to Fig. \ref{fig5},

\begin{figure}[h]
    \centering
   \includegraphics[width=0.6\textwidth]{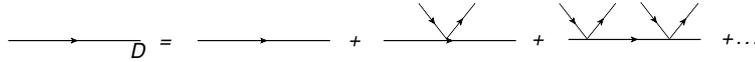}
    \caption{Dressed boson propagator}
    \label{fig4}
\end{figure}

 These graphs contribute with:
 
\be
i\int d^8\xi \int \frac{d^d k_E}{(2 \pi)^d}\frac{1}{\tilde{k}_E^2}\ln (1+\lambda\bar{\Phi}\Phi), 
\ee
\noindent
which gives zero in the context of  dimensional regularization. To determine the contribution corresponding to the  remaining diagrams, equations (\ref{ka}) and (\ref{kb}), we introduce the dressed $\Phi\bar{\Phi}$-propagator (see Fig. \ref{fig4}),

\be
\langle \Phi(\xi_1)\bar{\Phi}(\xi_2) \rangle_D=\frac{1}{1-\lambda\bar{\Phi}\Phi}\Pi_0\delta^8 (\xi_1-\xi_2).
\ee

By repeating a  calculation similar to the one  we did before we find finally:

\bea
\mathcal{K}_a&=&\int d^8\xi \int \frac{d^4 k_E}{(2 \pi)^4}\frac{1}{\tilde{k}_E^2}\left(\ln (1+\frac{g^2\bar{\Phi}\Phi}{\tilde{k}_E^2})-\ln (1+\frac{\lambda g^2\bar{\Phi}\Phi}{\tilde{k}_E^2})  \right) \\
\mathcal{K}_b&=&\int d^8 \xi \int \frac{d^4 k_E}{(2\pi)^4}\frac{1}{\tilde{k}_E^2} \ln\left(1+\frac{\lambda g^2\bar{\Phi}{\Phi}}{(1-\lambda\bar{\Phi}\Phi)(\tilde{k}_E^2+\lambda g^2\bar{\Phi}\Phi)}\right) .
\eea

In this case, the effective potential  is deformed by the factor $\frac{1}{1-\lambda\bar{\Phi}\Phi}$, but in the gauge where $\lambda\rightarrow 0$, the extra interaction $\int d^2\theta d^2\bar{\theta}\bar{\Phi}\Phi\bar{\Phi}\Phi$ does not affect the one-loop K\"ahler potential. The analysis of this situation is completely analog to the model we studied before.

\subsection{2-POINT FUNCTIONS: RESTAURATION OF THE LORENTZ SYMMETRY AT LOW ENERGY AND IMPROVEMENT OF THE UV BEHAVIOR }

In this section we study the one-loop 2-point functions of the scalar and vector superfields. We will see that at low energy, the standard Lorentz invariant SUSY Maxwell action is restored. For $z>1$ the two point function of the superfields are finite, improving the behavior of the Lorentz invariant case. For simplicity, we will work in the super-Fermi-Feynman gauge, $\lambda=1$. In this gauge the propagator of the vector superfield can be written as

\be
\langle V(\xi_1)V(\xi_2)\rangle=\frac{1}{\tilde{\square}}\delta^8(\xi_1-\xi_2)
\ee

\begin{figure}[h]
    \centering
   \includegraphics[width=0.6\textwidth]{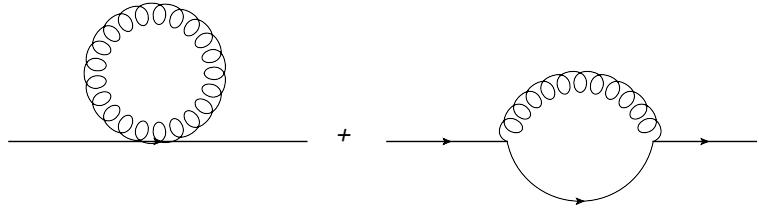}
    \caption{one-loop 2-point function, $\bar{\Phi}(-p)\Phi(p)$}
    \label{fig7}
\end{figure}

(Remember that the tilde notation refers to the deformed quantities) We start with the one-loop correction to the 2-point function of the scalar superfield $\Phi$ showed in Figure \ref{fig7}.  

\vspace{1cm}

The contribution of the tadpole diagram can be written as follows,

\be
\Sigma_2^{tadpole}(p)=-g^2\int d^4\theta_1d^4 \theta_2 \int\frac{d^4k}{(2\pi)^4}  \Phi(-\tilde{p},\theta_1)\bar{\Phi}(\tilde{p},\theta_2)\frac{1}{\tilde{k}^2}\delta_{12}\delta_{12},
\label{tadpolephi}
\ee
\noindent
which is zero since $\delta_{12}\delta_{12}=0$, while for the "fish" graph we have,

\be
\Sigma_2^{fish}(p)=-g^2\int d^4\theta \int \frac{d^4 k}{(2\pi)^4} \Phi(-\tilde{p},\theta)\bar{\Phi}(\tilde{p},\theta)A(\tilde{p},\tilde{k}),
\label{fishphi}
\ee
\noindent
where $A(\tilde{p},\tilde{k})=\frac{1}{\tilde{k}^2}\frac{1}{(\tilde{k}+\tilde{p})^2}$. 
The superfields in the expressions above were defined in equations (\ref{chiral}), and therefore we can expand them in momentum and truncate the expansion at second order. This means that we are disregarding high momentum contributions and hence the product of superfields $\bar{\Phi}(-p)\Phi(p)$ becomes Lorentz invariant. Let us call the truncated Lorentz invariant superfield $\Phi_{l.i.}(p)$, and then

\be
\bar{\Phi}(-\tilde{p})\Phi(\tilde{p})=\bar{\Phi}_{l.i.}(-p)\Phi_{l.i.}(p)+\mathcal{O}(p^z).
\ee

We will show that the function $A(\tilde{p},\tilde{k})$ is also Lorentz invariant for low $p$. First of all, we use the  Feynman parametrization to write the integral in the following way:

\be
\int \frac{d ^4k}{(2\pi)^4}\frac{1}{\tilde{k}^2}\frac{1}{(\tilde{k}+\tilde{p})^2}=\int_0^1 dx \int  \frac{d ^4k}{(2\pi)^4} \frac{1}{\left((\tilde{k}+x\tilde{p})^2-x^2 \tilde{p^2}+x\tilde{p}^2\right)^2}.
\ee

 We then shift the momentum $\tilde{k}$ as follows

\bea
&&k_0\rightarrow k_0-x \tilde{p}_0\\
&&k_i \left[c+\eta \left( \sum_{j=1}^3 k_i^2  \right)^\frac{z-1}{2} \right]\rightarrow k_i-x \tilde{p}_i.
\eea

Note that $\tilde{k}_i=k_i \left[c+\eta \left( \sum_{j=1}^3 k_i^2  \right)^\frac{z-1}{2}\right]$.
 We arrive finally at
 
 \be
 \int \frac{d^4 k }{(2\pi)^4} A(\tilde{p},\tilde{k})=\int _0^1 dx  \int \frac{d^4 k }{(2\pi)^4}  \vert J^{-1} \vert \frac{1}{\left(k^2-x^2\tilde{p}^2+x \tilde{p}^2\right)^2}\label{funA}
\ee

where $\vert J^{-1}\vert$ is the inverse Jacobian:

\be
\vert J \vert =c^3+c^2(2+z)\eta \vert\bar{k}^2\vert^\frac{z-1}{2}+c(1+2z )\vert\bar{k}^2\vert^\frac{z-1}{2}+z\eta^3 \vert\bar{k}^2\vert^\frac{3(z-1)}{2}.
\ee

Now, at low external momentum, after disregarding higher powers in $p$, we have $\tilde{p}^2=p^2+\mathcal{O}(p^z)$, being $p^2$ a Lorentz scalar. We arrive finally at

\be
\Sigma_2(p)=-g^2\int d^4\theta \bar{\Phi}(-p,\theta)_{l.i.}\Phi(p,\theta)_{l.i.}\int_0^1 dx\int \frac{d^4 k}{(2\pi)^4} \vert J^{-1} \vert \frac{1}{\left(k^2-x^2p^2+x p^2\right)^2}+\text{(high momentum)}\label{twopointphi}
\ee
\noindent

We observe that the two point function for $\Phi$ is Lorentz invariant at one loop order and therefore, in the low energy limit the usual quadratic Lorentz invariant action for the chiral superfield is generated. Moreover, for $z>1$ the integral in $k$ is convergent and the one loop two point funtion is finite.  We can follow the same procedure to obtain the two point function of the vector superfield$\,V$. 

\begin{figure}[h]
    \centering
   \includegraphics[width=0.6\textwidth]{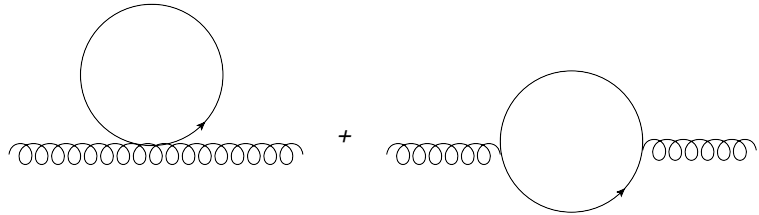}
    \caption{One-loop 2-point function, $V(-p)V(p)$}
    \label{fig8}
\end{figure}

The diagrams which contribute at one loop order are shown in Fig. \ref{fig8},

The contribution of the tadpole diagram can be written as follows,

\be
\Omega_2^{tadpole}(p)=g^2\int d^4\theta_1 d^4 \theta_2\frac{d^4k}{(2\pi)^4}  V(-p,\theta_1)\delta_{12}\frac{D^2\bar{D}^2}{16\tilde{k}^2}\delta_{12}V(p,\theta_2)
\label{tadpoleV}
\ee
\noindent
and for the ``fish" diagram,

\be
\Omega_2^{fish}(p)=\frac{g^2}{2}\int d^4\theta_1d^4\theta_2\frac{d^4k_E}{(2\pi)^4}  V(-p,\theta_1)\frac{\bar{D}^2 D^2}{16\tilde{k}^2}\delta_{12}\frac{D^2\bar{D}^2}{16(\tilde{k}+\tilde{p})^2}\delta_{12}V(p,\theta_2).\label{fishV}
\ee

Again we expand the vector superfield in powers of the momentum $p$, and disregard terms with powers greater than two. After that, the expansion of the term $ V(-p,\theta)V(p,\theta)$ becomes the Lorentz invariant one (we call the truncated superfield $V_{l.i.}$). After integration by parts (see for example \cite{YMEP3})  the tadpole contribution is compensated by part of (\ref{fishV}). The remaining contribution can be written as follows

\bea
\Omega_2(p)&=&\frac{g^2}{2}\int d^4\theta V_{l.i.}(-p,\theta) D^\alpha \bar{D}^2 D_\alpha V_{l.i.}(p,\theta)\times\nonumber\\ 
&&\times\int_0^1 dx\int \frac{d^4 k}{(2\pi)^4} \vert J^{-1} \vert \frac{1}{\left(k^2-x^2p^2+x p^2\right)^2}+\text{(high momentum)},
\label{2pointV}
\eea

\noindent
and it is, of course, Lorentz invariant. Therefore we have shown that in the low energy limit the Lorentz invariance is restored and the usual supersymmetric Maxwell action is generated.

\subsection{HIGHER ANISOTROPIC SPACE}

There is another interesting situation depending on the value of the critical exponent $z$. Let us analyze the behavior of the effective K\"ahler potential as $z$ increases. 
The intermediate cut off tends to $c$ and therefore the low energy contribution to the potential is always convergent, while the high energy contribution goes to $0$ (this can be deduced directly from expression (\ref{potK})). In the limit $z\rightarrow \infty$ we get,

\be
 \lim_{z\rightarrow \infty} \mathcal{K}(z,0)=\frac{\left(2 c^2+\mu ^2\right) \log \left(\sqrt{\frac{c^2}{\mu ^2}+1}+\frac{c}{\mu
   }\right)+c \left(c \log \left(\mu ^2\right)-\sqrt{c^2+\mu ^2}\right)}{8 \pi ^2 c^3}
\ee
\noindent
and hence in this limiting case, the K\"ahler potential can be calculated exactly and behaves as the low energy Lorentz invariant one. Moreover, we find a similar behavior for the two point functions of scalar and vector superfields. In this case, the integration over the internal momenta in the loop is performed over the compact region $[0,c]$,  the propagators are the Lorentz invariant ones (out of this region the contribution to the two point function is zero), and as a consequence all divergences dissapear. In general all n-amplitudes or the $z=\infty$ theory are determined by the usual lorentz invariant propagators, such that, every integration over internal momenta must be performed in the region $[0,c]$.

\section{Summary}

In this work we studied a Lorentz-violating superalgebra constructed with higher spatial derivative operators. With this modified  structure, we supersymmetrized a Lorentz-violating SUSY scalar QED theory. After the modification of the SUSY algebra  we were able to construct terms of the form $F_{0i}F_{0i}-F_{ij}\mathcal{O}F_{ij}$ (being $\mathcal{O}$ some function of the Laplacian  operator), although the superderivatives and supercharges do not verify the Leibniz rule. We showed how the UV behavior is improved, and we analyzed it with the explicit calculation of the superficial degree of divergence.

 We also obtained the one-loop  K\"ahler potential for arbitrary critical exponent $z$. We analyzed the UV behavior of the theory as well as the renormalization improvement and proved that for $z>1$ the effective K\"ahler potential is free of divergences. We studied different limits and showed how the theory naturally flows to the Lorentz invariant situation as $z$ approaches to one. We furthermore explore the possibility of more general interactions despite the fact that self-interactions of chiral and antichiral fields must be integrated in the whole Grassmann space and therefore contains derivative interactions in the component fields.  
 
 We analyzed the two point functions of the chiral and vector superfields and showed that, in the low energy limit, a Lorentz invariant supersymmetric QED is generated, for $z>1$ the two point functions of chiral and vector superfields being free of divergences.

{\bf Acknowledgements.} This work was partially supported by  Funda\c{c}\~{a}o de Amparo \`{a} Pesquisa do Estado de S\~{a}o Paulo (FAPESP) and Conselho
Nacional de Desenvolvimento Cient\'{\i}fico e Tecnol\'{o}gico (CNPq).

\appendix
\section{Two-point superfield Green's functions}

The quadratic action involving chiral superfields and chiral sources has the following form:

\be
S_{J,chiral}[\tilde{\Phi},\bar{\tilde{\Phi}},J,\bar{J}]=\int d^8 \xi\tilde{\Phi}\bar{\tilde{\Phi}}+\left(\int d^6\xi \tilde{\Phi} J+h.c\right),
\ee
\noindent
where all quantities are understood to be the tilde ones with derivatives and superfields defined in section II. We can write the previous action in terms of the measure $d^6\xi$ by using the  identity,

\be
\int d^8\xi  \tilde{\Phi}\bar{\tilde{\Phi}}=\left(\int d^6\xi \frac{1}{2}\tilde{\Phi} \left(-\frac{\bar{\tilde{D}}^2}{4}\right) \bar{\tilde{\Phi}}+h.c. \right),
\ee

which also holds for the usual Lorentz invariant SUSY. Therefore,

\be
S_{J,chiral}[\tilde{\Phi},\bar{\tilde{\Phi}},J,\bar{J}]=\int d^6 \xi_1 d^6 \xi_2 \begin{bmatrix} \tilde{\Phi}(\xi_1)&\bar{\tilde{\Phi}}(\xi_1) \end{bmatrix}\Delta \, \delta \begin{bmatrix} \tilde{\Phi}(\xi_2)\\ \bar{\tilde{\Phi}}(\xi_2)\end{bmatrix},
\ee

\noindent
where

\be
\Delta=
\left(
\begin{matrix}
0 & -\frac{1}{4}\bar{\tilde{D}}^2\\
-\frac{1}{4}\tilde{D}^2 & 0
\end{matrix}\right)
\ee

 and
 
\be
\delta=
\left(
\begin{matrix}
-\frac{1}{4}\tilde{D}^2\delta^8 (\xi_1-\xi_2) & 0 \\
0 & -\frac{1}{4}\bar{\tilde{D}}^2\delta^8 (\xi_1-\xi_2)
\end{matrix}\right).
\ee

Notice that everywhere appear the modified superderivatives. After integration over the chiral superfields we obtain the following partition function,

\be
Z[J,\bar{J}]=det ^{-1/2}\Delta \exp \{-\frac{i}{2}\int d^6 \xi_1 d^6 \xi_2\begin{bmatrix} J(\xi_1)&\bar{J}(\xi_1) \end{bmatrix}\Delta^{-1} \delta \begin{bmatrix} J(\xi_2)\\ \bar{J}(\xi_2)\end{bmatrix}   \},
\ee

\noindent
which allow us to determine the two-point Green function,

\be
G(\xi_1,\xi_2)=\frac{1}{i^2}\frac{\delta^2 Z[J,\bar{J}]}{\delta J(\xi_1)\delta \bar{J}(\xi_2)}=-i(-\frac{1}{4})^2 \frac{\bar{\tilde{D_1}}^2 \bar{\tilde{D_2}}^2}{\tilde{\square}}\delta^8(\xi_1-\xi_2).
\ee

By replacing the tilde quantities by the usual one this expression coincides with the known  Lorentz invariant expression.
 Similarly, we may obtain the vector propagators corresponding to (\ref{ss2}).


\begin{thebibliography}{99}
\bibitem{Stelle} K. S. Stelle, Phys. Rev. D 16, 953 (1977).
\bibitem{Hawking} S. W. Hawking, T. Hertog, Phys. Rev. D 65, 103515 (2002), hep-th/0107088.
\bibitem{antoniadis} I. Antoniadis, E. Dudas, D. M/ Ghilencea, Nucl. Phys. B767, 29 (2007), hep-th/0608094.
\bibitem{Lifshitz} E. M. Lifshitz, Zh. Eksp. Teor. Fiz. 11, 255 (1941)
\bibitem{anselmi} D. Anselmi, M. Halat, Phys.Rev. D76 (2007) 125011, arXiv:0707.2480.
\bibitem{Horava2} P. Horava, JHEP 0903 (2009) 020, arXiv : 0901.3775.
\bibitem{Horava} P. Horava, Phys.Rev. D79 (2009) 084008, arXiv : 0901.3775.
\bibitem{Iengo} R. Iengo, J. G. Russo, M. Serone, JHEP 0911,020 (2009), arXiv:0906.3477.
\bibitem{Pedro} P. R. S. Gomes, M. Gomes, Phys. Rev. D 85, 085018, (2012), arXiv:1107.6040.
\bibitem{Petrov6} M. Gomes , J. R. Nascimento, A. Yu. Petrov , A. J. da Silva, Phys. Rev. D 90 (2014) 12, 125022, arXiv:1408.6499.
\bibitem{redigolo} D. Redigolo, Phys.Rev. D85 (2012) 085009, arXiv:1106.2035.
\bibitem{Bagger} J. Wess and J. Bagger, SUSY and Supergravity. Princeton University Press, 1983.
\bibitem{Berger1} M.S. Berger, V.Alan Kostelecky, Phys.Rev. D65 (2002) 091701, hep-th/0112243
\bibitem{farakos1} K. Farakos, Int.J.Mod.Phys. A27 (2012) 1250168, arXiv:1204.5622 [hep-th
\bibitem{farakos2} K. Farakos, D. Metaxas, Phys.Lett. B711 (2012) 76-80, arXiv:1112.6080 [hep-th] 
\bibitem{farakos3} K. Farakos, D. Metaxas, Phys.Lett. B707 (2012) 562-565, arXiv:1109.0421 [hep-th]
\bibitem{Petrov5}  C. F. Farias, J. R. Nascimento, A. Yu. Petrov, Phys.Lett. B719 (2013) 196-199, arXiv:1208.3427 [hep-th]
\bibitem{Petrov8} M. Gomes, T. Mariz, J. R. Nascimento, A. Yu. Petrov, J. M. Queiruga and A. J. da Silva, 	
On one-loop corrections in the Horava-Lifshitz-like QED. arXiv:1504.04506. 
\bibitem{YMEP1} B. de Wit, M. T. Grisaru, M. Rocek, Phys. Lett. B  374, 297 (1996), hep-th/9601115
  \bibitem{YMEP2} A. Pickering, P. West, Phys. Lett. B  383, 54 (1996), hep-th/9604147
  \bibitem{YMEP3} M. T. Grisaru,
  M. Rocek, R. von Unge, Phys. Lett. B  383, 415 (1996), hep-th/9605149
\bibitem{Petrov7} A.C. Lehum , J.R. Nascimento, A. Yu. Petrov , A.J. da Silva, Phys.Rev. D88 (2013) 045022, arXiv:1305.1812 [hep-th]
\bibitem{ovrutgal} Justin Khoury, Jean-Luc Lehners , Burt A. Ovrut, Phys.Rev. D84 (2011) 043521,  arXiv:1103.0003 [hep-th]
\bibitem{ovrutghost} Justin Khoury, Jean-Luc Lehners , Burt A. Ovrut, Phys.Rev. D83 (2011) 125031, arXiv:1012.3748 [hep-th] 
\bibitem{kehagias} Fotis Farakos, Alex Kehagias, PoS Corfu2012 (2013) 127, arXiv:1305.1784 [hep-th]
\bibitem{queiruga} C. Adam, J.M. Queiruga, J. Sanchez-Guillen, A. Wereszczynski, JHEP 1305 (2013) 108, arXiv:1304.0774 [hep-th] 








\end{thebibliography}
\end{document}